\documentclass[12pt]{article}
\usepackage{latexsym}
\usepackage{graphicx}
\usepackage[dvipsnames]{xcolor}
\usepackage{caption}
\usepackage{psfrag}
\usepackage{amsmath,amssymb,dsfont}
\newcommand{\be}{\begin{equation}}
\newcommand{\ee}{\end{equation}}
\newcommand{\bea}{\begin{eqnarray}}
\newcommand{\eea}{\end{eqnarray}}
\newcommand{\lbl}{\label}

\newcommand{\nn}{\nonumber}
\newcommand{\re}{\mathrm{e}}

\newcommand{\borel}{\mathcal{B}^{[\rho]}(\sigma)}
\newcommand{\borelPT}{\mathcal{B}^{[\rho]}_{\rm PT}(\sigma)}
\newcommand{\bl}{\mathcal{B}_{<}^{[\rho]}(\sigma)}
\newcommand{\bh}{\mathcal{B}_{\mathrm{>}}^{[\rho]}(\sigma)}

\newcommand{\s}{\sigma}
\newcommand{\sh}{\widehat{\sigma}}

\newcommand{\dF}{d^{(F)}}

\newcommand{\ie}{{\it i.e.}}

\definecolor{darkgreen}{rgb}{0.0, 0.4, 0.0}

\long \def \blockcomment #1\endcomment{}

\input epsf

\numberwithin{equation}{section}

\begin{document}

\begin{titlepage}

\begin{flushright}
\end{flushright}
\vspace*{1.5cm}
\begin{center}
{\Large \bf Hyperasymptotics and quark-hadron duality violations in QCD}\\[1.5cm]

Diogo Boito,$^a$ Irinel Caprini,$^b$
Maarten Golterman,$^{c,d}$
Kim Maltman,$^{e,f}$ Santiago Peris$^c$
\\[8mm]
{\small\it
$^a$Instituto de F\'isica de S{\~a}o Carlos, Universidade de S{\~a}o Paulo,
CP 369\\ 13570-970, S{\~a}o Carlos, SP, Brazil
\\[5mm]
$^b$Horia Hulubei National Institute for Physics and Nuclear Engineering, POB MG-6\\
077125 Bucharest-Magurele, Romania
\\[5mm]
$^c$Department of Physics and IFAE-BIST, Universitat Aut\`onoma de Barcelona\\
E-08193 Bellaterra, Barcelona, Spain
\\[5mm]
$^d$Department of Physics and Astronomy,
San Francisco State University\\ San Francisco, CA 94132, USA
\\[5mm]
$^e$Department of Mathematics and Statistics,
York University\\  Toronto, ON Canada M3J~1P3
\\[5mm]
$^f$CSSM, University of Adelaide, Adelaide, SA~5005 Australia}
\end{center}

\vspace*{1.0cm}

\begin{abstract}
We investigate the origin of the quark-hadron
duality-violating terms in the expansion of the QCD two-point vector
correlation function at large energies in the complex  $q^2$ plane.
Starting from the dispersive representation for the associated
polarization, the analytic continuation of the operator product
expansion from the Euclidean to the Minkowski region is performed by
means of a generalized Borel-Laplace transform, borrowing techniques
from hyperasymptotics. We establish a connection between singularities
in the Borel plane and quark-hadron duality violating contributions.
Starting with the assumption that for QCD at $N_c=\infty$ the
spectrum approaches a Regge trajectory at large energy, we obtain an
expression for quark-hadron duality violations at large, but finite $N_c$.
\end{abstract}

\end{titlepage}

\begin{boldmath}
\section{Introduction}
\lbl{sec0}
\end{boldmath}
Correlation functions in QCD, at sufficiently large energy, can be
calculated starting from the gluon and quark degrees of freedom,
using perturbation theory, augmented by the operator product
expansion (OPE). Many of these correlators, such as the Adler
function of the vector current two-point function, can also be expressed,
through dispersion relations,
in terms of experimentally accessible spectral functions.  These
spectral functions reveal the presence of multiple hadronic
resonances, whose spectral contributions are not reproduced
when perturbative and higher-dimension OPE contributions are evaluated
on the Minkowski axis.

In spite of this difference, a complete
description in terms of Lagrangian or physical degrees of
freedom should be equivalent, a notion which is referred to as quark-hadron
duality. It is, however, generally accepted that even at
large energies, resonance effects, and hence contributions beyond
the OPE, are present in QCD correlators in the Minkowski
region.\footnote{In this paper, we will consider the purely
perturbative contribution as the leading term in the OPE.}
These additional contributions, which by definition violate
quark-hadron duality, are usually referred to as duality
violations (DVs). Their origin, and possible models for their
form, have been the subject of many earlier papers
\cite{PQW}-\cite{Peris}, but a more formal derivation of their
form has not been achieved thus far. It is clear that DVs have
to exist, as the OPE is, at best, an asymptotic series. This is
intuitively obvious from the fact that the imaginary part of
the OPE for the Adler function does not look anything like the
physical spectral function, except for asymptotically large energies.

In this paper, we present a more systematic investigation of
the form DVs may take, limiting ourselves to the case of the
Adler function for simplicity. Since DVs are a consequence
of the appearance of resonances in the spectrum,
the properties of the resonance spectrum must represent
a starting point for our analysis. Of course, very little
is known analytically about spectral functions beyond
perturbation theory. But, if we can show, starting from
a general and physically motivated assumption about the form
of the resonance spectrum, that this assumption is compatible
with the known form of the OPE for large Euclidean momenta,
we expect the form of DVs for Minkowski momenta implied by
this same assumption to also represent a good approximation to
the form of DVs in QCD.\footnote{We note
that the matching of an assumed form of the resonance spectrum
in large-$N_c$ to the OPE has been considered before in
Ref.~\cite{Pineda}.}

We begin by working in the limit $N_c\to\infty$
(where $N_c$ is the number of colors), where the spectral
function is known to be given by an infinite sum of
Dirac $\delta$-functions, consistent with asymptotic
freedom \cite{Wittenbaryons}. We will make assumptions
about the form of the resonance spectrum in this limit
that lead us to the known form of the OPE for Euclidean
momenta, and then use this information to derive the form
of the associated duality-violating contributions to the Adler
function on the Minkowski axis. It turns out that with some
additional assumptions, this analysis can then be generalized
to large, but finite $N_c$. For simplicity, we will work
in the chiral limit, so that the Adler function depends only on one
variable, which can be taken to be the ratio of the
momentum to the QCD scale.

Technically, our task will be to analytically continue the Adler
function in the complex $q^2$ (momentum-squared) plane from the
Euclidean to the Minkowski region. Our starting point is an
assumption for the form of the spectral function, which defines
the Adler function through a dispersion relation. It turns out
to be advantageous to reformulate the problem as one where we
write the Adler function $\mathcal{A}(q^2)$ as a Borel-Laplace
transform involving a new function, $\borel$, which is itself the
Laplace transform of the spectral function. In
hyperasymptotics \cite{Math}, the appearance of duality-violating
terms, \ie, terms beyond the OPE, as a result of
analytic continuation in the Borel plane is understood in terms
of the concept of a ``transseries," for which the OPE represents
the first term, with exponential corrections \cite{AU,MStrans}.
We will see how singularities in the Borel plane lead to various
terms in the transseries, with singularities at the origin
corresponding to the OPE, and those at non-zero distance
from the origin to higher-order transseries terms. In fact,
the OPE itself can be viewed as a transseries which goes
beyond perturbation theory, and the singularities in the Borel
plane associated with perturbation theory are nothing else
than the well-known renormalons \cite{tHooft}-\cite{ Beneke}.
Higher-order terms in the OPE appear as the effect of renormalon
singularities in the Borel plane.\footnote{We will recover this
result in the course of our study of the Adler function in this
paper.} Much less is known about the non-perturbative singularities,
but it is clear that their physical origin is in the non-perturbative
physics of the spectral function. It follows that we will need
physical input, which will be provided by means of a rather general
assumption about the form of the resonance spectrum, in combination
with the analytic continuation in the Borel plane, to arrive at an
analytic form for the duality-violating contributions to the Adler function.

This paper is organized as follows. In the next section, we
give the representation of the Adler function in
terms of a Borel-Laplace transform, and  show how this
representation can be used to analytically continue in the
complex $q^2$ plane, starting from the Euclidean region
Re~$q^2<0$, emphasizing the essential role played by the
singularities of the Borel transform in the complex Borel
plane. The subsequent sections investigate the
Borel-plane singularities in a sequence of models of gradually
increasing complexity and, at the same time, of increasing
resemblance to QCD as well.

We begin, in Sec.~\ref{sec2}, with a simple Regge model for the
spectrum in the large-$N_c$ limit. This allows us to demonstrate
how singularities at the origin in the Borel plane correspond
to the OPE, while DVs are associated with singularities away
from the origin. Then, in Sec.~\ref{sec1}, we generalize our
{\it ansatz} for the spectrum in the large-$N_c$ limit to a much
more general form. In Secs.~\ref{OPE} and \ref{sec1prime} we
show how we may recover perturbation theory and the OPE in
the ``large-$\beta_0$'' approximation in which all except
the first coefficient of the $\beta$-function are set equal
to zero. In particular, in Sec.~\ref{sec1prime} we discuss
how the pure perturbative series and the singularities it
generates in the Borel plane, which are relatively well
understood, fit into the general picture. This discussion
also allows us to rederive the well-known SVZ sum rules \cite{SVZ}.
In Sec.~\ref{nonpert} we show how the appearance of DVs
from singularities away from the origin in the Borel plane
generalizes from the simple model of Sec.~\ref{sec2}. In
particular, we show that the singularities away from the
origin in the Borel plane stay in the same location, but
change from simple poles to branch points.

We then expand the discussion to large, but finite $N_c$.
The new ingredient is, of course, that now the hadronic resonances
become unstable. As we will see, and as has been observed
previously, now the duality-violating corrections become
exponentially suppressed, as observed in nature. We first show
how this works in the simple Regge model in Sec.~\ref{sec3prime},
before treating the more general case in Sec.~\ref{sec4}, in
which we arrive at our main result. Sec.~\ref{sec:conclusions}
contains our conclusions, while a number of technical details
have been relegated to Apps.~\ref{dirichletapp} to~\ref{app1}.
In App.~\ref{sec:numerics}, we compare, numerically, results
for the values of the DV parameters obtained from analyses of hadronic
$\tau$-decay data in Ref.~\cite{Boito3} with those obtained
from the fits to Regge trajectories of Ref.~\cite{Masjuan},
finding remarkable agreement.

\vskip0.4cm
\begin{boldmath}
\section{Borel-Laplace transform}
\lbl{sec:bl}
\end{boldmath}
We recall that the Adler function is defined as\footnote{The Adler
function is sometimes denoted by $D(q^2)$. We choose a normalization
such that it is equal to one at leading order in perturbation
theory.}
\be\lbl{eq:A}
\mathcal{A}(q^2)=-q^2\frac{d}{dq^2}\Pi(q^2)\ ,
\ee
where $\Pi(q^2)$ is the scalar correlator of the vector
current two-point function. From causality
and unitarity, we know that $\Pi(q^2)$ is an analytic function of
real type, \ie, it satisfies the Schwarz reflection principle
$\Pi((q^2)^*)=[\Pi(q^2)]^*$ in the complex $q^2$ plane cut along
the real axis above the lowest hadronic threshold, $4 m_\pi^2$.
The known asymptotic behavior of $\Pi(q^2)$ ensures that it
can be represented by a once-subtracted dispersion
relation,
\be\label{eq:DR}
\Pi(q^2)=\Pi(0)+q^2 \int_0^\infty \frac{\rho(t)}
{t(t-q^2-i\epsilon)} dt\ ,
\ee
in terms of the spectral function
\be\label{eq:rho}
\rho(t)=\frac{1}{\pi}\, {\rm Im}\, \Pi(t+i\epsilon)\ .
\ee

The polarization, $\Pi (q^2)$, and the Adler function depend on the
single variable, $q^2$. In practical applications of QCD, this
dependence is encapsulated in two different series expansions.
Taking $q^2<0$ Euclidean, one series is written in powers of
$\alpha_s(q^2)$ and the other in inverse powers of $q^2$ itself,
\be
\lbl{eq:QCD}
\mathcal{A}(q^2)_{\rm OPE}=1+ \sum_{n\ge 1} c_n\, \alpha^n_s(-q^2)
+\sum_{n\ge 1} \frac{d_n(q^2)}{(-q^2)^n}\ ,
\ee
where the first and second terms correspond to the perturbative
series in powers of the running strong coupling $\alpha_s(q^2)$
and the third term may be associated with the condensate expansion
of the OPE (the coefficients $d_n(q^2)$ depend logarithmically on $q^2$).
The corresponding expansion of $\Pi(q^2)$ is obtained in a
straightforward way using Eq.~(\ref{eq:A}). The interplay between
the two series in Eq.~(\ref{eq:QCD}) is at the origin of the difficulties encountered
in QCD phenomenology when trying to assess the relative importance
of perturbative {\it vs.} nonperturbative contributions.

At one loop the dependence of the strong coupling $\alpha_s(q^2)$
on $q^2$ is given by
\be\label{eq:a1loop}
  \alpha_s(-q^2)=\frac{1}{\beta_0\log (-q^2/\Lambda^2)}\ ,
\ee
where $\Lambda^2$ is the QCD parameter after the renormalization
scheme is specified, and $-\beta_0<0$ is the first coefficient of
the $\beta$ function.  At higher orders, the coupling exhibits a
more complicated logarithmic dependence on $q^2$ which, in fact,
depends on the precise definition chosen for this coupling.

Both series in Eq.~(\ref{eq:QCD}) are divergent, and in each case,
one expects corrections to the series exponential in the inverse
of the expansion parameter. Indeed, the power corrections in
Eq.~(\ref{eq:QCD}) can be interpreted in this way as corrections
to the perturbative series, since, with $\mbox{exp}[-1/
[\beta_0\alpha_s(-q^2)]] =\Lambda^2/(-q^2)$, inverse
powers of $-q^2$ are exponential in the inverse of the
strong coupling.

The objective of this paper is to see if, starting from a reasonable
form for the physical spectral function $\rho(t)$, \ie, a spectral
function that is physically sensible, and from which one recovers
the structure of Eq.~(\ref{eq:QCD}) for Euclidean $q^2$, one can
find the corrections to the OPE for $q^2>0$, \ie, in the Minkowski
region. Again, we expect these corrections to be exponential in
the inverse of $\Lambda^2/q^2$, possibly modified by logarithms.
This would allow us to make contact between the OPE representation
for the spectral function, obtained by analytically continuing
the expansion~(\ref{eq:QCD}) to the Minkowski region, and
the additional contributions from DVs.

Combining Eqs.~(\ref{eq:A}) and~(\ref{eq:DR}), the dispersive
representation for the Adler function can be written as a
Borel-Laplace transform,
 \bea
 \lbl{adler}
 \mathcal{A}(q^2)
 &=& -\, q^2\int_0^{\infty}dt\ \rho(t) \int_0^{\infty}
d\sigma \ \sigma\ \mathrm{e}^{-\sigma\left(t-q^2\right)}\\
 &=& -\, q^2 \int_0^{\infty} d\sigma\ \mathrm{e}^{\sigma q^2}
\, \sigma \borel\ ,\nn
 \eea
where
 \be
 \lbl{rhohat}
  \borel=\int_0^{\infty}dt\  \rho(t)\  \mathrm{e}^{-\sigma t}
  \ee
is the Laplace transform of the spectral function. We note that
$\borel$ is well-defined for any $\sigma$ with Re~$\sigma>0$,
since $\rho(t)$ goes to a constant for $t\to\infty$. Any
singularities of $\borel$ thus have to reside in the
half-plane Re~$\sigma\le 0$. This representation of the
Adler function in terms of $\borel$ is valid for Re$\,q^2<0$.
It follows that
  \be
  \lbl{intadler}
  \Pi(q^2)= C+\int_0^{\infty} d\sigma\ \mathrm{e}^{\sigma q^2}\, \borel\ ,
  \ee
with $C$ a regularization-dependent constant.

Provided $q^2<0$, \ie, in the Euclidean regime, it is clear that a
series expansion in powers of $\sigma$ of the function
$\sigma\borel$ translates into an asymptotic expansion
of $\mathcal{A}(q^2)$ in powers of $1/q^2$ that we may associate with the
OPE.\footnote{These powers are modified
by logarithmic terms; for instance, a $\log\sigma$ term in $\borel$
will generate a $\log(-q^2)$ correction. Such $\log(-q^2)$ corrections are
screened by at least one power of
$\alpha_s$ in the Adler function.} In particular, the function $\sigma \borel$ must go to a
constant as $\sigma \rightarrow 0$ for $\mathcal{A}(q^2)$ to
reproduce the parton-model constant in the limit
$-q^2\rightarrow \infty$, which is the first term in the
$1/q^2$ expansion. In general, the OPE is expected to be asymptotic.

Because the OPE is not an expansion with a finite radius of
convergence, it cannot directly be used for $\mathrm{Re}\, q^2>0$,
and in particular not on the Minkowski axis. As we will see, the
analytic continuation to the Minkowski axis will produce new
contributions, the duality-violating terms. These corrections
are defined as the difference between the exact Adler function
and its quark-gluon representation in terms of the OPE,
for large energies. The central question we attempt to address
here is: what is the form of these  corrections to the OPE
when we analytically continue from the Euclidean axis,
$q^2<0$, to the Minkowski axis, $q^2>0$?

\begin{figure}
\begin{center}

\leftline{\includegraphics*[width=3cm]{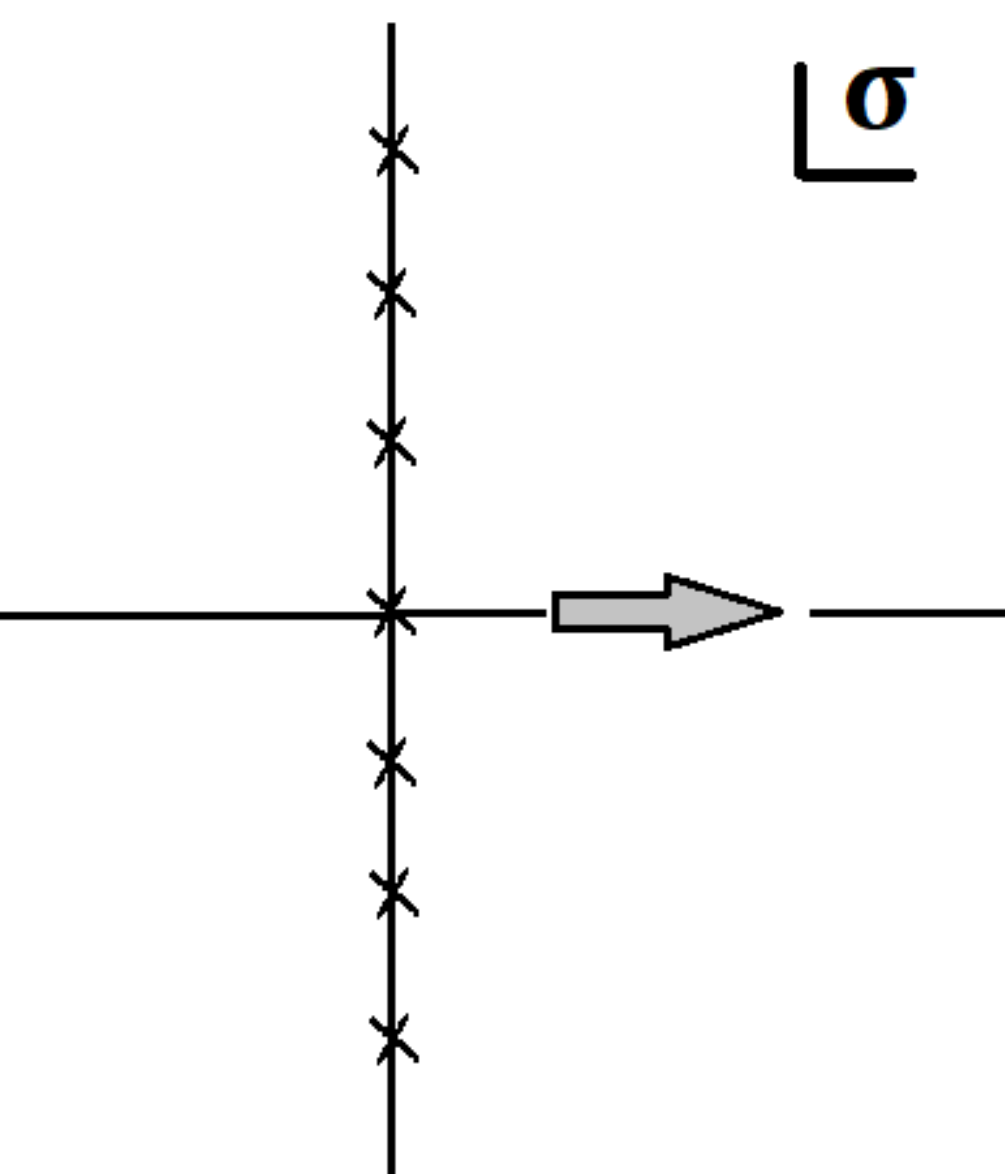}}

\vspace{-2cm}$ \hspace{2cm} \text{\Huge $\Leftrightarrow$} \hspace{2cm}$

\vspace{-2.5cm}\rightline{ \includegraphics*[width=4cm]{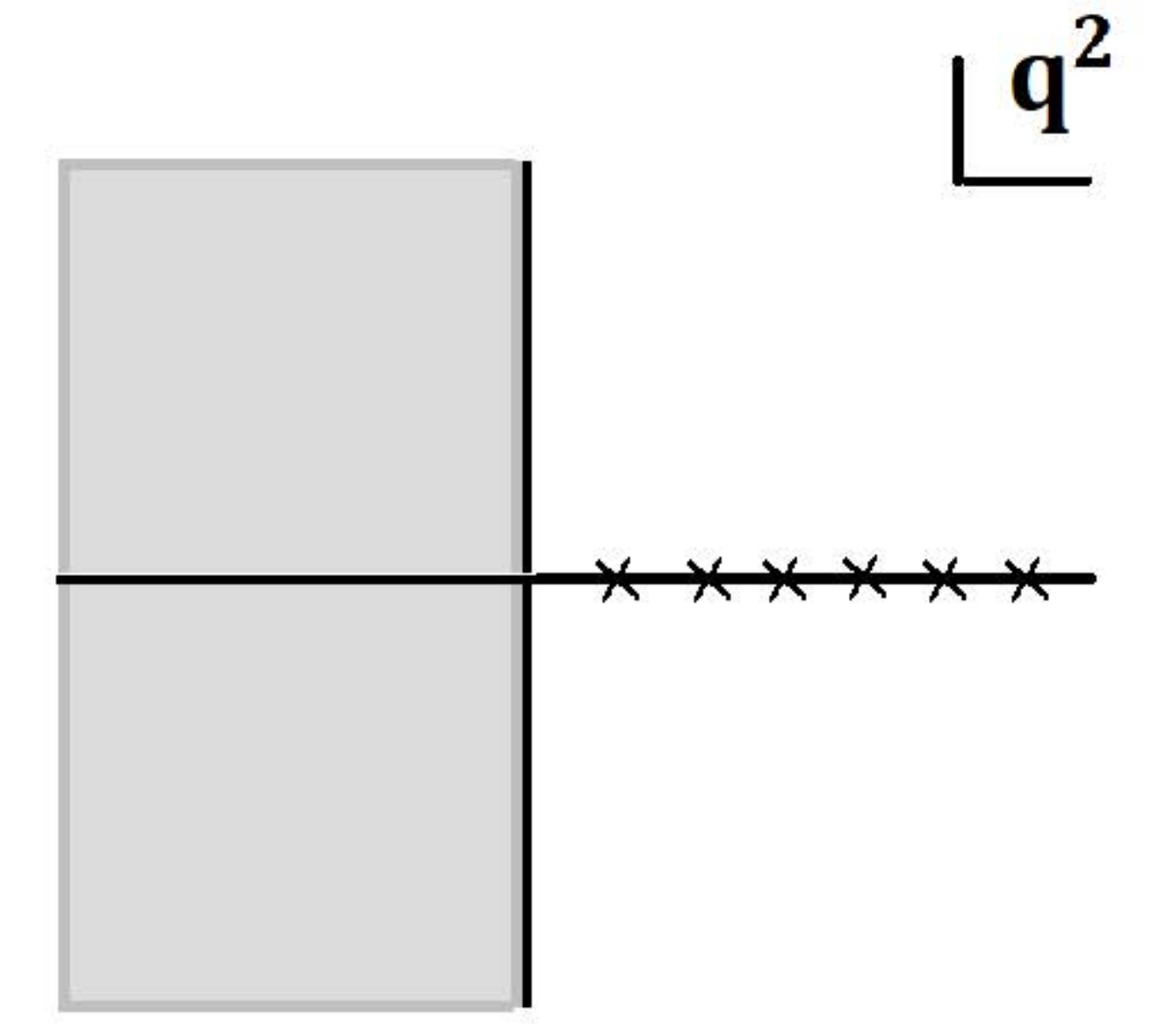}}

\vspace{.5cm}

\leftline{\hspace{-.3cm}\includegraphics*[width=4cm]{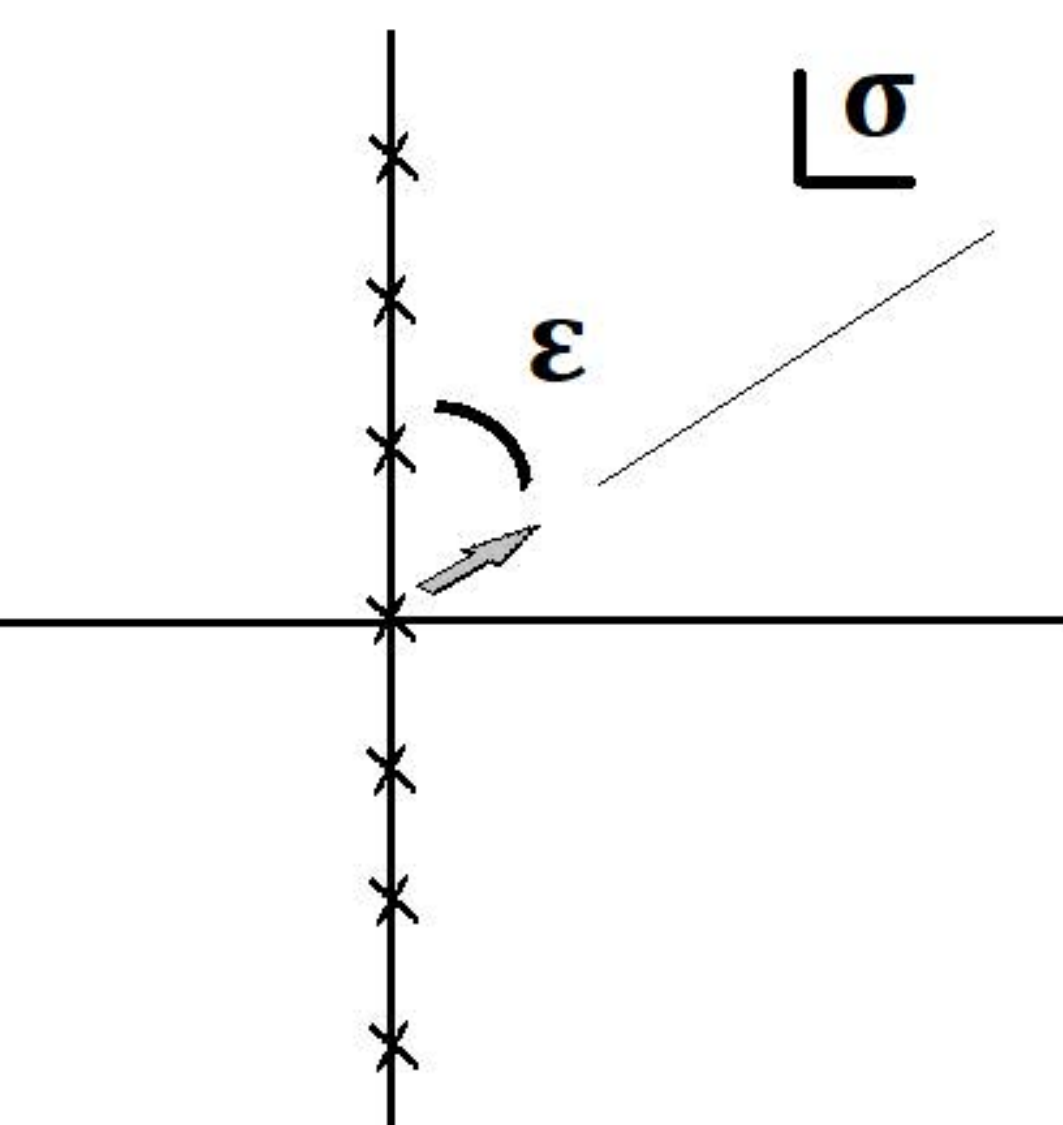}}

\vspace{-2cm}$ \hspace{2cm}  \text{ \Huge $\Leftrightarrow$}\hspace{2cm}$

\vspace{-2.5cm}\rightline{\includegraphics*[width=4cm]{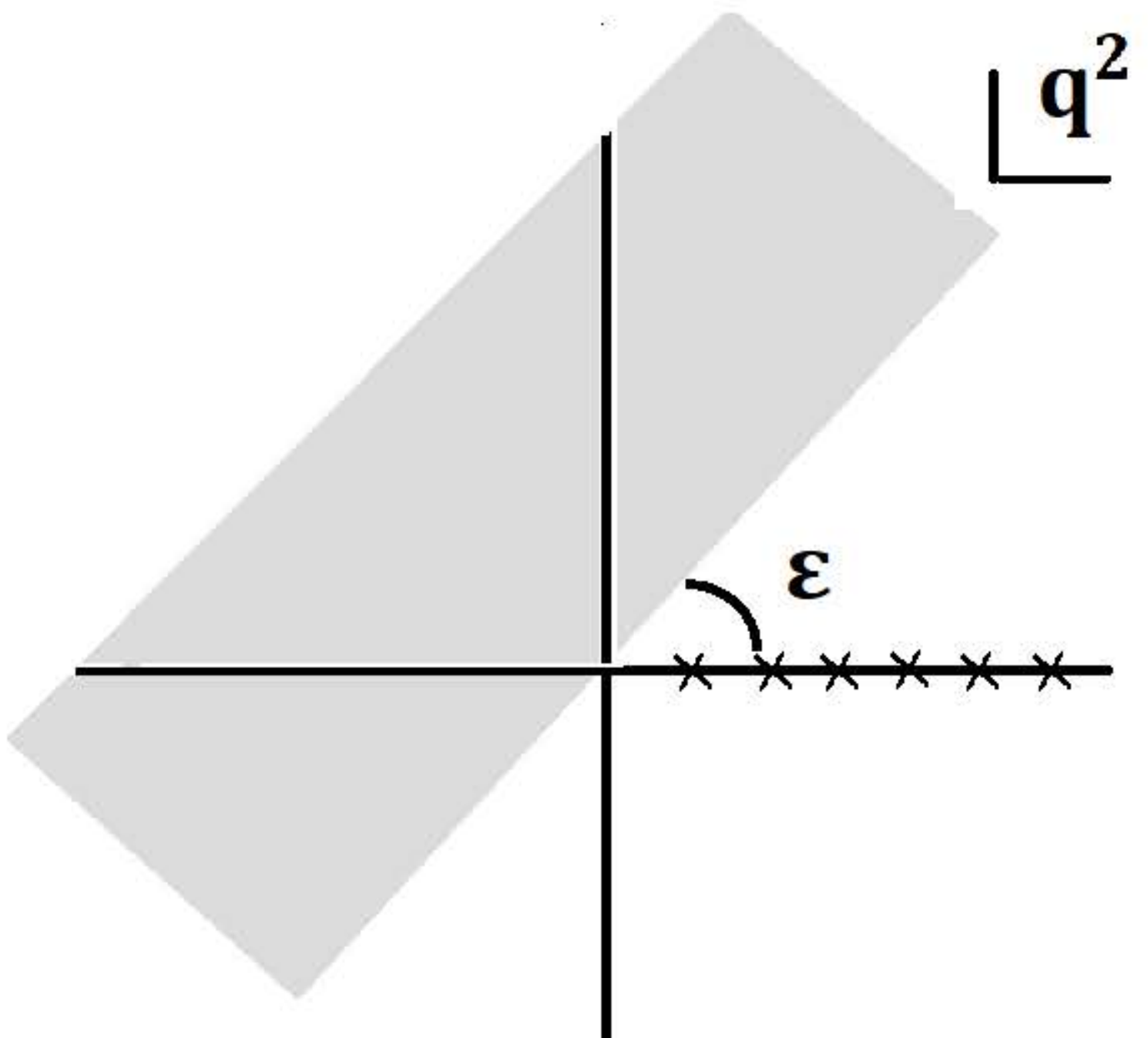}}

\vspace{1cm}

\leftline{\hspace{-.9cm}\includegraphics*[width=4cm]{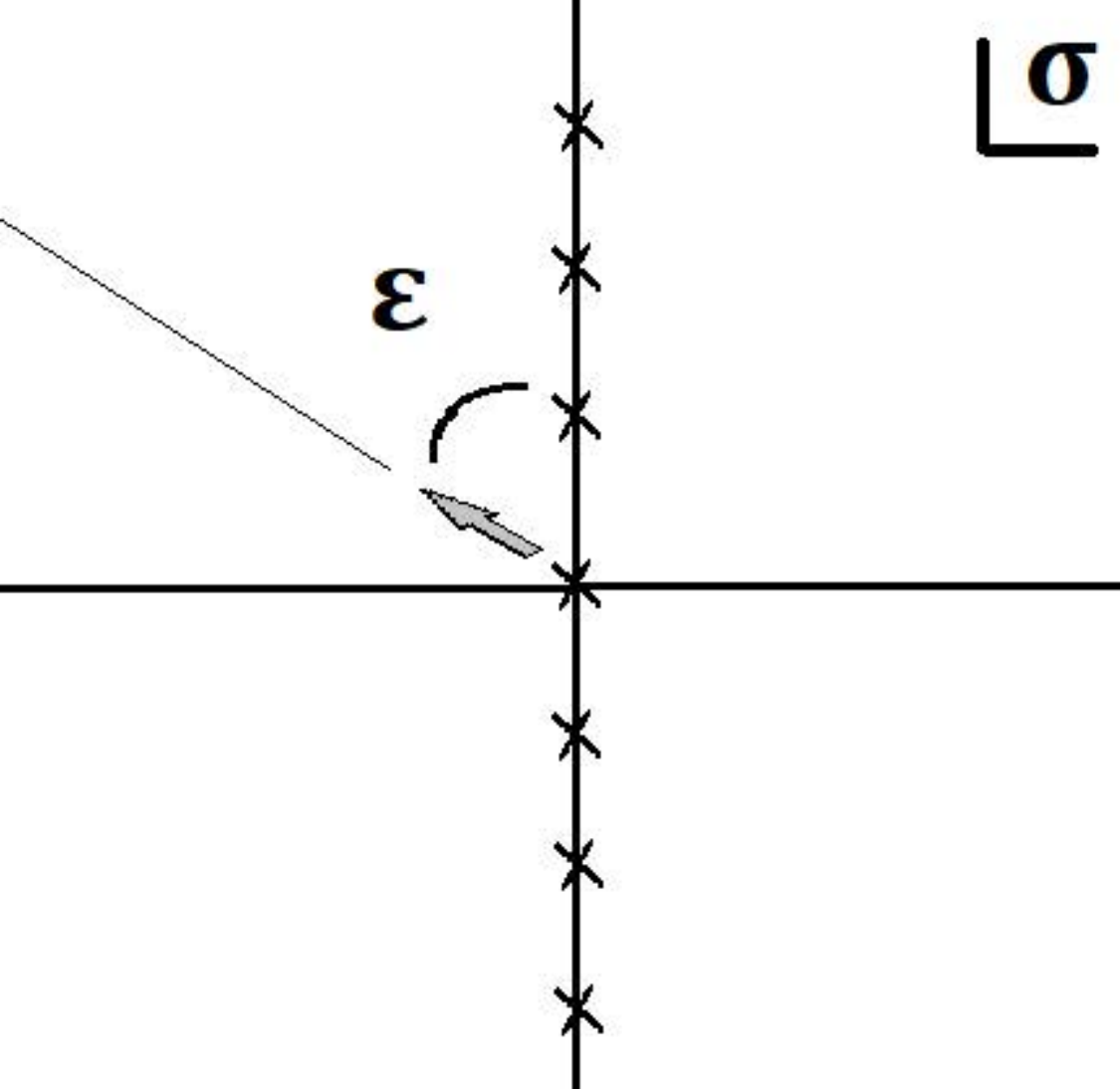}}

\vspace{-2cm}$ \hspace{2cm}  \text{ \Huge $\Leftrightarrow$}\hspace{2cm}$

\vspace{-3cm}\rightline{\includegraphics*[width=4cm]{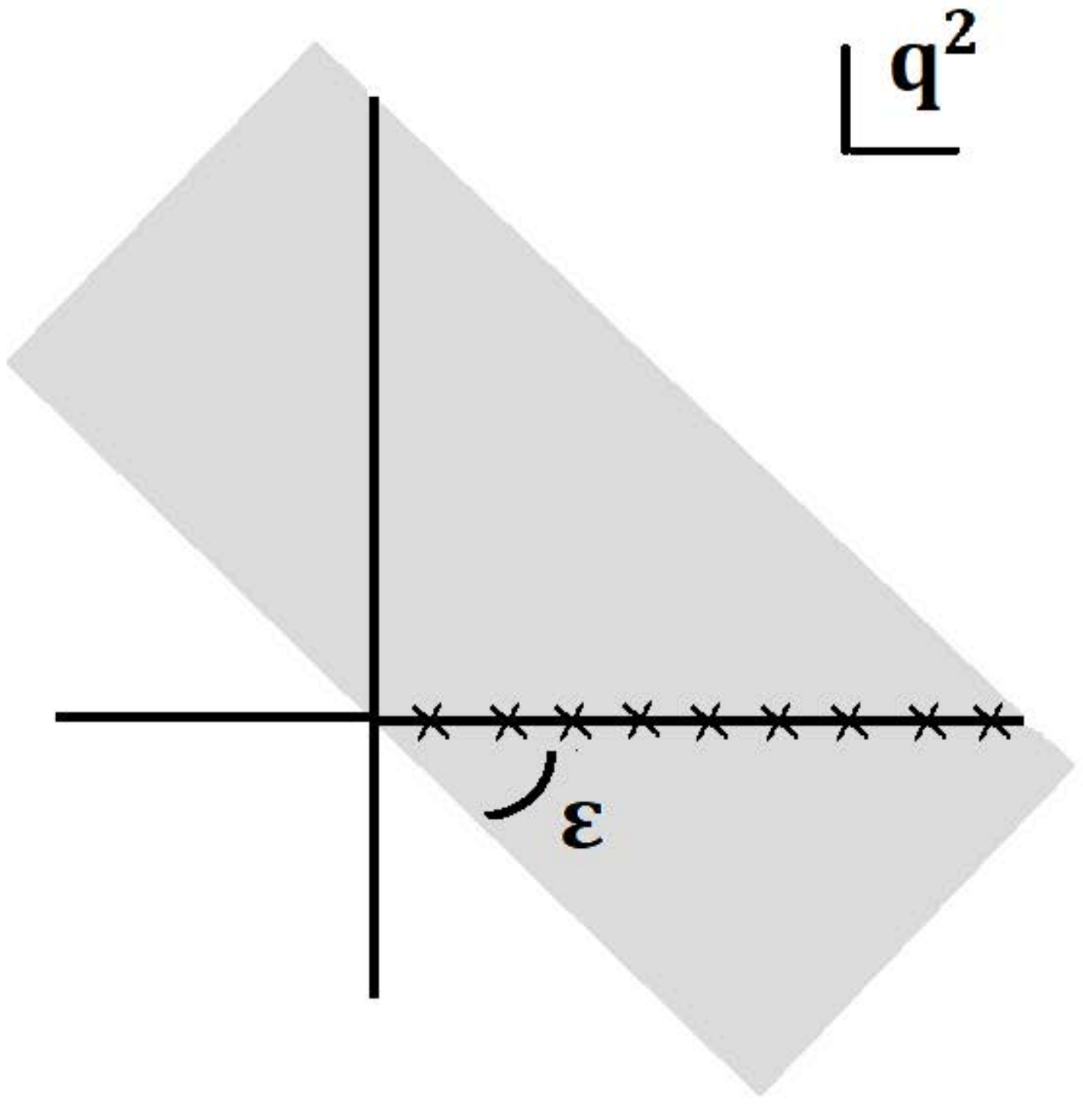}}

\vspace{1cm}

\leftline{\hspace{-1cm}\includegraphics*[width=4cm]{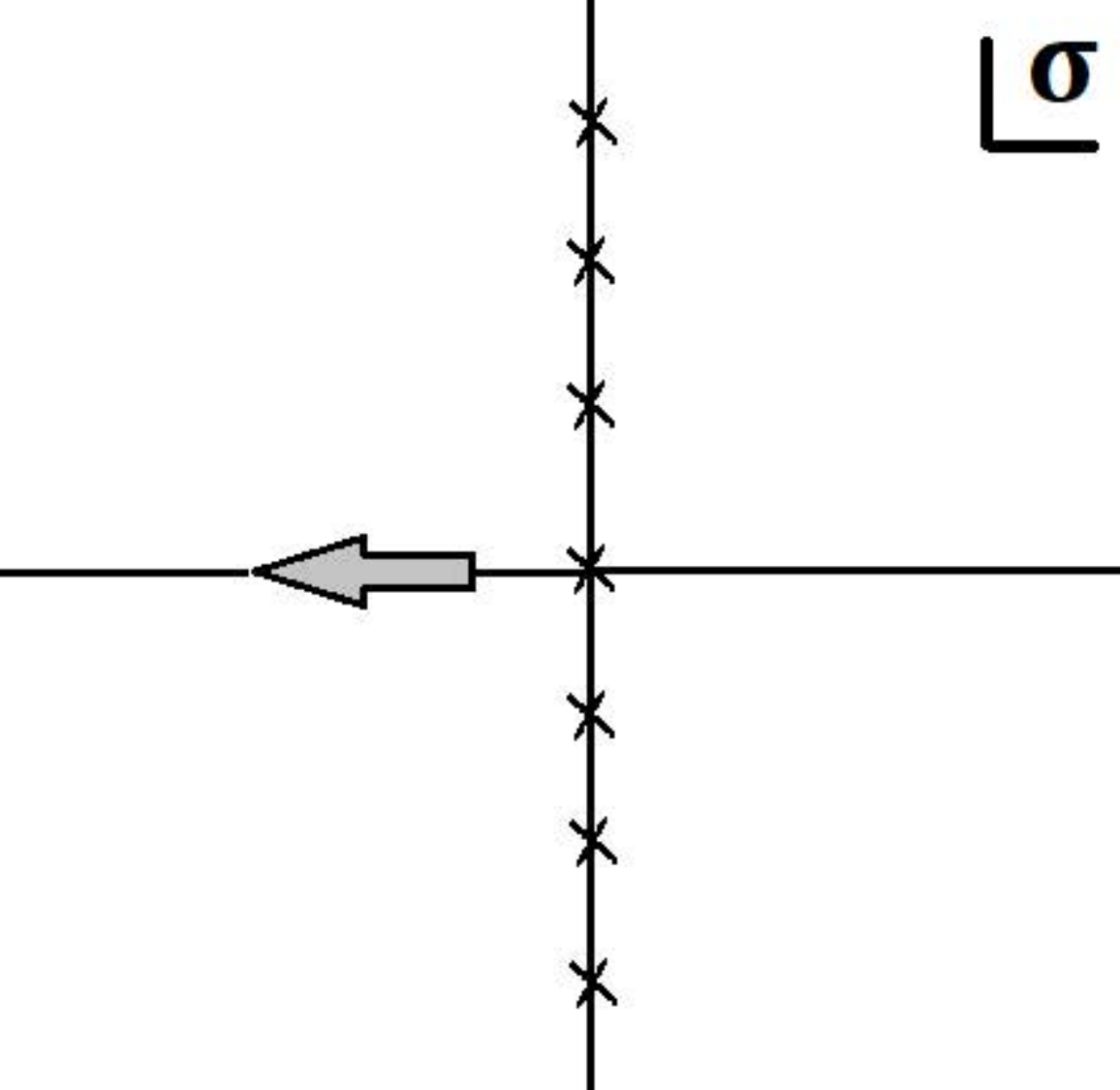}}

\vspace{-2.2cm} $ \hspace{2cm}  \text{ \Huge $\Leftrightarrow$}\hspace{2cm}$

\vspace{-2.5cm}\hspace{-1cm}\rightline{\includegraphics*[width=3cm]{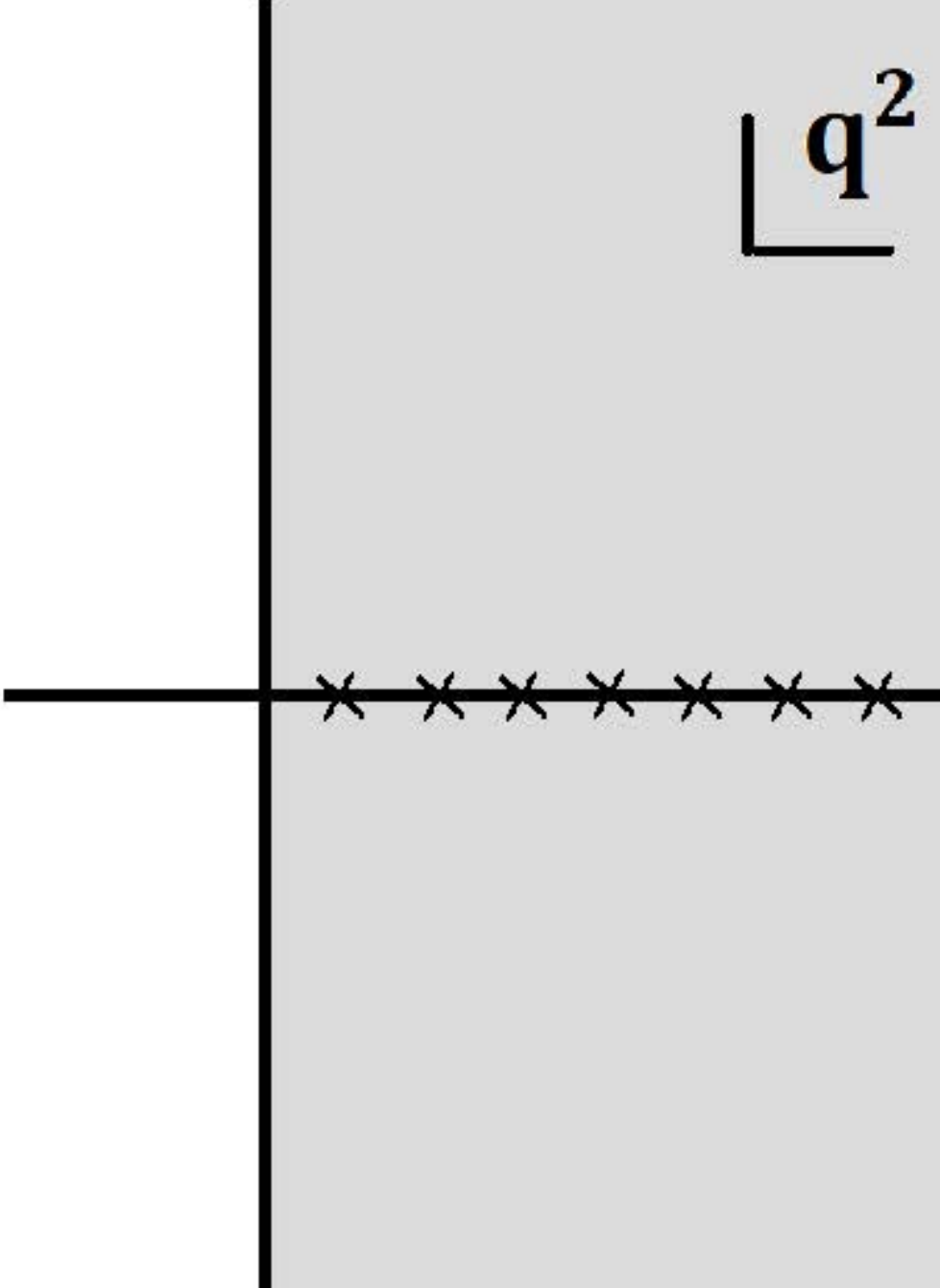}}

\includegraphics*[width=4cm]{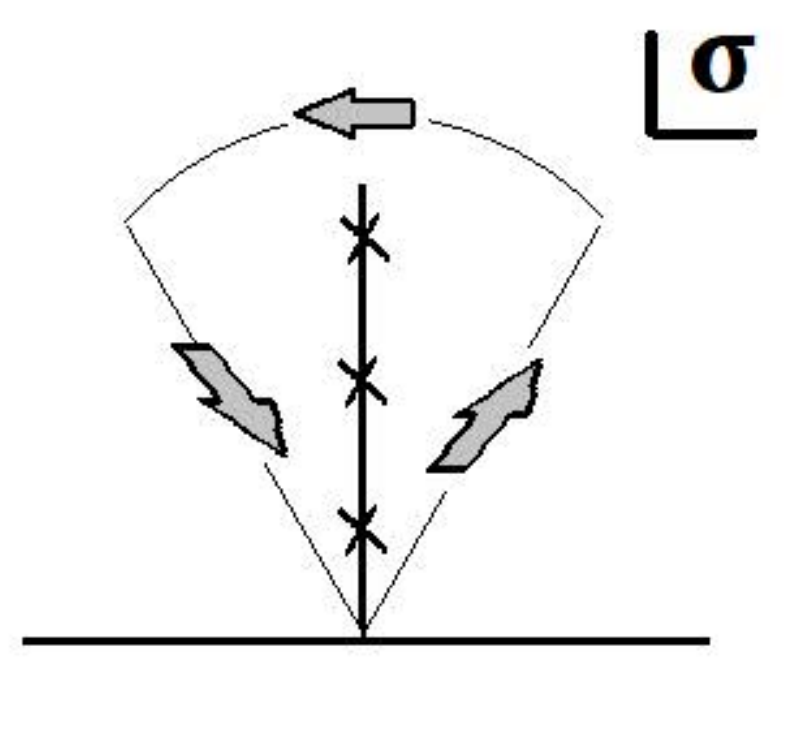}
\end{center}
\caption{Analytic extension using the generalized Borel-Laplace
transform, $\borel$, in the large-$N_c$ limit.
Crosses denote poles or branch points in the $\sigma$ plane and
the associated poles in the spectrum in the $q^2$ plane.}
\end{figure}

The integral~(\ref{adler}) has the form of a Borel-Laplace
transform and is defined for $\sigma$ along the positive real
axis and for $\frac{\pi}{2}< \arg q^2< \frac{3\pi}{2}$,
\ie, for $\mathrm{Re}\,q^2<0$.\footnote{We will define the
first Riemann sheet of the $q^2$ complex plane as
$q^2=|q^2| \re^{i \varphi},\ 0\leq \varphi <2 \pi$.
For later use, we will define the sheet with $-2\pi\leq\varphi<0$
as the zeroth Riemann sheet, {\it etc.}} However, this definition
can be generalized by considering different rays (all starting at
the origin) in the complex $\sigma$ plane defined by varying the
angle $\arg\sigma$, as long as
$\frac{\pi}{2}< \arg \sigma + \arg q^2 <\frac{3\pi}{2}$
so that the integral in Eq.~(\ref{adler}) remains well
defined. By varying $\arg\sigma$ the generalized
Borel-Laplace transform thus defined analytically extends the
definition of $\mathcal{A}(q^2)$ to larger regions in the
$q^2$ complex plane \cite{Math}.

With $\arg \sigma=\frac{\pi}{2}-\epsilon\ (\epsilon>0)$, for example,
the region in $q^2$ covered becomes $\epsilon< \arg q^2 < \pi + \epsilon$.
Note that this region partly overlaps with the the region we started with,
$\frac{\pi}{2}< \arg q^2< \frac{3\pi}{2}$, as an analytic
continuation should do. If no singularities in the $\sigma$
plane are crossed as $\arg \sigma$ rotates from $0$ to
$\frac{\pi}{2}-\epsilon$, and if the function  $\sigma\, \borel$
does not grow exponentially on the contour at infinity
connecting these two angles, the result of the two integrals will
be the same in the region of overlap. In order for this analytic
continuation to work, we have to assume that
$e^{\sigma q^2}\sigma\, \borel$ decays to zero at the
circle at infinity for $|q^2|$ arbitrarily large. We believe
that this assumption is satisfied by the representations for
$\borel$ considered in this paper.

To analytically extend the Borel-Laplace integral~(\ref{adler})
to the Minkowski axis, it is thus essential to know the location and
nature of the singularities of the function $\sigma\, \borel$ in the
complex $\sigma$ plane.  As an example, in Fig. 1 we have
depicted the singularities in the $\s$ plane we will encounter in
the large-$N_c$ example of Sec.~\ref{sec2}. As $\sigma$ is
rotated from the positive to the negative real axis,
anticlockwise, for example, the presence of any such
singularity will add a contribution to the analytic
continuation of $\mathcal{A}(q^2)$ for $q^2>0$. Such
extra contributions are the source of the duality-violating
contributions to $\mathcal{A}(q^2)$.

Of course, instead of rotating anticlockwise, we could also
choose to rotate clockwise in the complex $\sigma$ plane. If
we always choose the anticlockwise rotation, the resulting
Adler function would not satisfy the Schwarz reflection property,
$\mathcal{A}((q^2)^*)=[\mathcal{A}(q^2)]^*$. However, we can enforce
this property by limiting the anticlockwise rotation to $q^2$ with
Im~$q^2\ge 0$, while, instead, rotating clockwise for values of
$q^2$ with Im~$q^2< 0$. Thus, for negative values of Im~$q^2$,
the rotation at the heart of our analytic continuation should
be changed into a clockwise rotation in order to enforce the
reflection property. In the rest of this paper, we will always
use the anticlockwise rotation, and obtain the spectral function
from the Adler function at $q^2+i\epsilon$ with $q^2$ real
and positive. Generally, $\mathcal{A}(q^2)$ for Im~$q^2<0$ can
be obtained from our results through the Schwarz
reflection relation.

It is clear from Eq.~(\ref{rhohat}) that the singularity
structure of $\sigma\borel$ is directly determined by
the spectrum and that, with present
technology, it is not possible to calculate this singularity
structure from first principles in the case of QCD.
However, there are important qualitative aspects of the
spectrum generally assumed to be properties of QCD and,
as we will see, these are sufficient to infer with some
confidence what type of singularities in $\sigma\borel$ one may
expect. Furthermore, these general observations can be backed up
with explicit calculations in model examples, as we will
demonstrate below. In the next section, we will consider
a concrete example, in order to illustrate the mechanism
described qualitatively above. This concrete example will
then serve as a starting point for a much more general
discussion in Sec.~\ref{sec1}, in which we will make contact
between the assumed form of the resonance spectrum and the
OPE, before using this form to deduce the functional
dependence of DVs on $q^2$.

\vskip0.4cm
\begin{boldmath}
\section{Example: A simple Regge model for $N_c=\infty$}
\lbl{sec2}
\end{boldmath}
To simplify the discussion, we first consider the large-$N_c$
limit, leaving the generalization to finite $N_c$ to Secs.~\ref{sec3prime}
and \ref{sec4}. In this limit, we know that the full set of
singularities of $\Pi (q^2)$ in the complex $q^2$ plane consists
of an infinite sequence of simple poles located at ever increasing
values of $q^2$ on the positive real $q^2$ axis. The spectral function is
the corresponding sum of Dirac $\delta$ function contributions. As $\sigma$
is rotated from $\arg \sigma=0$ to $\arg \sigma=\pi/2+\epsilon$
the region of validity in $q^2$ of the representation (\ref{adler})
shifts from $\pi/2< \arg q^2 <3\pi/2$ to $-\epsilon< \arg q^2 <\pi-\epsilon$,
the latter encompassing all the poles of the spectrum.
It is clear that these singularities prevent one from performing
a naive analytic continuation in $q^2$ of the Adler
function originally defined for $q^2<0$ in Eq.~(\ref{adler}). Since $q^2$
touches the positive real axis when $\sigma$ touches the positive
imaginary axis, some singularities must exist for $\arg \sigma=\pi/2$
which reflect the singularities for $\arg q^2=0$.

To see in detail what is going on, we now consider the example
of a model in which the spectral function is given by an infinite
set of delta functions on a linear trajectory, \ie,
\be
\lbl{rhotoy}
\rho(t)=\sum_{n=1}^{\infty} F(n)\delta\!\left(t-M^2(n)\right)\quad,
\quad  n=1,2,3,\dots\ ,
\ee
with
\be
\lbl{massdecay}
M^2(n)=\Lambda^2 n\  , \qquad F(n)=F^2\ .
\ee
We will use units such that $\Lambda=1$, and rescale the spectral function
such that $F=1$. This spectrum is not arbitrary. It corresponds to the
leading Regge behavior in an expansion for large $n$, where $n$ is the
resonance excitation number. In two-dimensions, this asymptotic Regge
behavior represents the actual spectrum of QCD in the  large-$N_c$ limit
\cite{tHooft2d,CCG} and, although to date it has never been proved,
asymptotic Regge behavior is generally believed to be true in
large$-N_c$ QCD also in four dimensions. The string
picture \cite{string} and phenomenology \cite{Masjuan}
also provide some evidence for this behavior.

Using this spectrum, the function $\sigma \borel$ is easily shown
to be
\bea
\lbl{rhohattoy}
\sigma \borel= \frac{\sigma}{\re^{\sigma}-1}
=\sum_{n=0}^\infty \frac{B(n)}{n!}\,\sigma^n
 \  ,
\eea
where $B(0)=1, B(1)=-1/2$ and
$B(n> 1)=\frac{(-1)^{n+1} 2 (2n)!}{(2\pi)^{2n}}\, \zeta(2n)$
are the Bernoulli numbers and $\zeta(s)$ is the Riemann $\zeta$-function.
The function in Eq.~(\ref{rhohattoy}) has simple poles at $\sigma=\pm  2k\pi i\  (k=1,2,3,\dots)$,
with residues $\pm  2k\pi i$ ({\it cf.} the crosses in the left panels
of Fig. 1; the cross at $\sigma=0$ is removed by the factor $\sigma$).

As we try to extend the definition of $\mathcal{A}(q^2)$ to real
$q^2>0$ by rotating $\sigma$, we of course hit these poles at
$\arg\sigma=\frac{\pi}{2}$. As before, having increased $\arg\sigma$
from $0$ to $\frac{\pi}{2}-\epsilon$, the region of validity of
Eq.~(\ref{adler}) has shifted to $\epsilon< \arg q^2<\pi+\epsilon$. The
poles of the spectrum at $n=1,2,3,\dots$ seen in Eq.~(\ref{rhotoy})
(Fig. 1, second to top panel, right) now lie just outside this new region.
Clearly, there is a correspondence between these singularities
and the singularities of the Borel function $\sigma\borel$. If the
correlator had no singularities on the positive real $q^2$
axis it would be possible to analytically continue $\mathcal{A}(q^2)$
to include this axis, \ie, to move from the region
$\epsilon\leq\arg q^2<\pi+\epsilon$ to the region
$-\epsilon<\arg q^2\leq\pi-\epsilon$. However, the correlator
does have poles on the positive real axis, and this is also
reflected in the $\sigma$ plane: as $q^2$ crosses the positive
real axis, $\sigma$ crosses the positive imaginary axis on
which the poles of $\sigma\borel$ are located (Fig. 1, third
to top panel).

Letting $\sigma$ cross the imaginary axis, \ie, going from the
second to top to the third to top panels in Fig. 1 and reaching
$\arg\sigma=\frac{\pi}{2}+\epsilon$ produces a change in the
Borel-Laplace integral because now closing the contour at
infinity between the two rays encircles the singularities on
the positive imaginary $\sigma$ axis (Fig. 1, single panel at
the bottom).\footnote{The integral over the relevant portion of the
circle at infinity vanishes.}
No further singularities are encountered, and hence
no further contributions generated, as $\arg \sigma$
is rotated from $\frac{\pi}{2}+\epsilon$ to $\pi$ and,
with $\Gamma$ the contour depicted at the bottom of Fig.~1, one obtains
\be
\lbl{adlerminkowski}
\frac{d\Pi}{dq^2}(q^2)=\int_{\arg\sigma=\pi}d\sigma\
\re^{\sigma q^2} \sigma\borel+\frac{d\Pi_{\rm DV}}{dq^2}(q^2)\ ,
\quad (q^2>0)\ ,
\ee
where the ``duality violating'' contribution
$\frac{d\Pi_{\rm DV}}{dq^2}(q^2)$ has been defined as
\be
\lbl{adlerDV}
\frac{d\Pi_{\rm DV}}{dq^2}(q^2)=\int_{\Gamma} d\sigma\
\re^{\sigma q^2}\sigma\borel\  .
\ee
A straightforward use of Cauchy's theorem  leads to
\be
\lbl{adlerminkowskitoy}
\frac{d\Pi_{\rm DV}}{dq^2}(q^2)=2 i\pi
\frac{d}{dq^2}\sum_{k=1}^{\infty} \re^{i 2 k\pi q^2}
=-\pi \frac{d}{dq^2}\left(\cot \pi q^2 + i \right)\ .
\ee
We remark that this result does not satisfy the Schwarz reflection property.
However, it is straightforward to check that if we use
Eq.~(\ref{adlerminkowskitoy}) to define $\Pi_{\rm DV}$ for
Im~$q^2\geq 0$, but instead carry out a clockwise rotation
for the analytic continutation to the half-plane Im~$q^2<0$,
the resulting definition of $\Pi_{\rm DV}$ does satisfy the
reflection property.

Integrating Eq.\,(\ref{rhohattoy}), one obtains
\be
\lbl{dpsi}
\int_{\arg\sigma=\pi}d\sigma\ \re^{\sigma q^2}\sigma\borel
= - \frac{d\psi}{dq^2}(q^2)\ ,
\ee
where $\psi(q^2)=d\log{\Gamma(q^2)/dq^2}$ is the digamma function.
Integrating Eq.~(\ref{adlerminkowski})  with respect to $q^2$ one thus finds
\be
\lbl{pitoy}
\Pi(q^2)=-\psi(q^2)-\pi \left(\cot \pi q^2 + i \right)+ c\quad ,\quad (q^2>0)
\ee
where $c$ is an integration constant which can be fixed by the
condition $\mathrm{Im}\,\Pi(0)=0$ to be $i\pi$ plus an
undetermined real part.\footnote{ $\Pi(0)$ is a real constant
which depends on the renormalization scheme.}
We emphasize the emergence of the cotangent function in the
process of analytic continuation; we did {\it not} invoke the symmetry
property
\be
\lbl{psifct}
\psi(z)=\psi(-z)- \pi \cot(\pi z)-\frac{1}{z}\ ,
\ee
as was done in previous discussions of this model \cite{Blok,Cata1,Cata2}.
In other words, the process of analytic continuation allows us to rederive
this global property of the $\psi$ function.

Use of the representation
\be
\lbl{cot}
-\pi \cot\pi q^2= -\frac{1}{q^2}- 2 q^2 \sum_{n=1}^{\infty}
\frac{1}{(q^2+n)(q^2-n)}\ ,
\ee
immediately leads to
\be
\lbl{checkspectrum}
\frac{1}{\pi}\,\mathrm{Im}\,\Pi(q^2+i\epsilon)
=\sum_{n=1}^{\infty} \delta(q^2-n)\ , \quad (q^2>0) \ ,
\ee
which reproduces the initial spectrum, as it should ({\it cf.}
Eq. (\ref{rhotoy})). We see that the information about the spectrum is
contained in the DV term of Eq.~(\ref{adlerminkowski}). We will
now also show that the first term in Eq.~(\ref{adlerminkowski})
is indeed the analytic continuation to $q^2>0$ of the OPE series
obtained from Eq.~(\ref{adler}), with $q^2<0$.

Denoting by $[q^{2n}]\mathcal{A}_{\rm OPE}(q^2)$ the $(q^2)^{-n}$
term in the $1/q^2$ expansion of $\mathcal{A}(q^2)$ and using
the representation of Eq.~(\ref{adler}), one obtains
\be
\lbl{OPE1}
[q^{2n}]\mathcal{A}_{\rm OPE}\left(q^2<0\right)=-q^2\
\frac{B(n)}{n!}\ \int_{\arg \sigma=0}\!\!\!
d\sigma\ \mathrm{e}^{\sigma q^2}\sigma^n =
\frac{B(n)}{(-q^2)^{n}}\int_0^{\infty}
\frac{dt}{n!}\ \mathrm{e}^{-t} t^n= \frac{B(n)}{(-q^2)^{n}}\ ,
\ee
where the change of variables $t=-q^2\sigma>0$ has been made.
For $q^2>0$, instead, the representation to be used is the
one of Eq. (\ref{dpsi}), and one obtains
\bea
\lbl{OPE2}
[q^{2n}]\mathcal{A}_{\rm OPE}\left(q^2>0\right)
&=&-q^2\ \frac{B(n)}{n!}\ \int_{\arg \sigma=\pi}\!\!\!\!\!
d\sigma\ \mathrm{e}^{\sigma q^2}\sigma^n\ , \\
 & =&-q^2\ \frac{B(n)}{n!}\  \int_{0}^{-\infty}\!\!\!\!\!
d\sigma\ \mathrm{e}^{\sigma q^2} \sigma^n=  \frac{B(n)}{(-q^2)^{n}n!}
\int_0^{\infty}\!\! dt \, \mathrm{e}^{-t}\, t^n=\frac{B(n)}{(-q^2)^{n}}\nn \ .
\eea
Clearly the two results are the same. The factorial behavior of $B(n)$ implies that
the expansion is asymptotic.

We end this section with two comments.  The first is that, even in the
Euclidean regime, Re~$q^2<0$, the series~(\ref{OPE1}) is asymptotic.
This is a consequence of the fact that the Borel transform, $\sigma\borel$, has a
finite radius of convergence equal to $2\pi$, the distance of the
singularity closest to the origin in the Borel plane.
The OPE is obtained by expanding $\sigma\borel$ around $\sigma=0$ and
inserting this expansion into Eq~(\ref{adler}). Since the series
in $\s$ does not converge in the full integration interval, this
yields an asymptotic series for the OPE as $-q^2\to \infty$.
If one cuts off the integral at $\sigma=2\pi$, it is straightforward
to show that, at any finite order in the OPE, the remainder is
of order $\mbox{exp}(-2\pi |q^2|)$ \cite{Cata1}.
We see that, in this model, the presence of the singularities at
$\sigma=\pm 2\pi i$ has two consequences. First, it affects the nature
of the OPE for $q^2<0$, where the integral in Eq.~(\ref{adler}) is
well defined, and, second, it affects the analytic continuation
to the Minkowski regime, leading to the DV contribution to the Adler
function shown in Eq.~(\ref{adlerminkowski}).

The second comment is that no logarithmic terms are present in the
asymptotic expansion of the simple model (\ref{rhotoy}). As we will see
in the following sections, such terms arise only if large-$n$ subleading
corrections are added to the model. This simple example,
however, demonstrates how the properties of the spectrum
are reflected in the singularities of the function $\sigma\borel$,
which, in turn, determine the form of the DVs.
We note that the singularities in $\sigma\borel$ which correspond
to the duality violating part of $\Pi(q^2)$ lie on the
imaginary $\sigma$ axis. We will argue in the next section that
with subasymptotic corrections to Regge behavior, but still in the
large-$N_c$ limit, these singularities will stay on the
imaginary axis, though they will no longer be simple poles.
It then follows that, if large-$N_c$ is a good approximation,
they will have to stay close to the imaginary axis for finite
$N_c$, and thus will remain well separated from the cuts in
the $\sigma$ plane along the negative real axis which correspond to
the perturbative corrections to the OPE.

\vskip0.4cm
\begin{boldmath}
\section{A generalized Regge spectrum for large-$N_c$ QCD}
\lbl{sec1}
\end{boldmath}

In the previous section we saw the consequences of assuming a
linear trajectory for the spectrum. In  this approximation,
$\sigma\borel$ has simple poles on the positive imaginary axis and
the OPE it generates contains no logarithms, but only powers
of $1/q^2$.  How does the picture change when terms which are
subleading at large resonance excitation number $n$ are
also taken into account? As we will now see, subleading terms
change the nature of the singularities from simple poles
to branch points, without modifying their location, and
introduce logarithmic corrections into the series in
powers of $1/q^2$.

In general, the function $\borel$, for Re$\,\s>0$, is given
in large-$N_c$ QCD by a series of the form
\be
 \lbl{dirichlet}
 \borel=\sum_{n=1}^\infty F(n)\, \re^{-\sigma M^2(n)}\ ,
\ee
where the $F(n)$ can in general be complex numbers and $\{M^2(n)\}$
is a monotonically  increasing sequence of non-negative real numbers
tending to infinity (\ie, there is no accumulation point). For the
particular case of the vector-current polarization considered here,
the quantities $F(n)$ are
real and positive. Series like Eq.~(\ref{dirichlet}) are known as
Dirichlet series \cite{Dirichlet}.

The concepts of a radius, boundary and disk of convergence in a
power series are replaced for a Dirichlet series by the abscissa,
line and half-plane of convergence. The line of convergence is
the value $\sigma=\sigma_c$ such that for
$\mathrm{Re}\, \sigma >\sigma_c$ the Dirichlet series converges
while for $\mathrm{Re}\, \sigma <\sigma_c$ it diverges. The
region $\mathrm{Re}\, \sigma >\sigma_c$ is called the half-plane
of convergence. In our case the line of convergence will be
located at Re$\,\sigma=0$, \ie, the imaginary axis.  To our
knowledge, the first article to point out that Dirichlet
series are relevant to the study of large-$N_c$ QCD was
Ref. \cite{deRafael}.\footnote{This reference speculates about the  connection
between a complex pole in the $\sigma$ plane and the origin of
DVs, on the basis of a purely mathematical
model.}

Inspired by two-dimensional QCD, we will assume the spectrum
to obey the following expansion at large $n$ \cite{FLZ}:
\bea
\lbl{reggespectrum}
F(n)&=&1+\epsilon_F(n)\  ,\\
M^2(n)&=&n+ b\log n+ c+ \epsilon_M(n)\ ,\nn
\eea
where $b$ and $c$ are constants, and
\bea
\lbl{corrections}
\epsilon_{i}(n)&=&\epsilon_{i}(0,n)+ \epsilon_{i}(\{\lambda\},n)
\ ,\quad  i=F, M\quad  ,\\
\epsilon_{i}(0,n)&=&\sum_{\nu_{i}>0}\frac{d^{(i)}(\nu_{i})}
{(\log n)^{\nu_{i}}}\ ,
\nonumber\\
\epsilon_{i}(\{\lambda\},n)&=&
\sum_{\lambda_{i}> 0,\nu_{i}}\frac{d^{(i)}(\lambda_{i},\nu_{i})}
{n^{\lambda_{i}} (\log n)^{\nu_{i}}}\ .
\nonumber
\eea
We take the values of $\lambda_i$ in these expressions, and those of
the $\nu_i$ in $\epsilon_{i}(0,n)$, to be positive,
while the values of $\nu_i$ in $\epsilon_{i}(\{\lambda\},n)$
are allowed to be positive, negative or zero. The
$\epsilon_{i}(n)$ are subleading contributions in the sense that
$\epsilon_{F}(n)\rightarrow 0 $ and $\epsilon_{M}(n)/n\rightarrow 0$
as $n\to \infty$. The correction $\epsilon_{i}(0,n)$ in
Eq.~(\ref{corrections}) is, in fact, just a special case
of $\epsilon_{i}(\{\lambda\},n)$ with $\lambda =0$. We choose
to split it off because it turns out to generate the logarithms
that appear in perturbation theory, whereas the
$\epsilon_{i}(\{\lambda\},n)$ corrections with non-zero $\lambda$
contribute to the power corrections. We do not know whether subleading
terms of a different form may occur in large-$N_c$ QCD,
but the forms assumed in Eq.~(\ref{reggespectrum}) turn out to
be sufficient for our purpose, which is to further investigate
the relation of the detailed structure of the spectrum to
the OPE. The behavior $F(n)\to 1$ as $n\to \infty$
(up to an overall multiplicative
constant) is a consequence of the asymptotic Regge
spectrum $M^2(n) \sim n$, as $n\to \infty$, and a requirement to
obtain the leading-order parton-model result at large $-q^2$.

The rest of this section consists of three parts. First, in
Sec.~\ref{OPE}, we obtain the OPE  of the Adler function at
large Euclidean momenta from Eq.~(\ref{reggespectrum}). In
Sec.~\ref{sec1prime} we {focus, in particular, on the first term
in the OPE, \ie, perturbation theory. Then, in Sec.~\ref{nonpert},
we generalize the discussion of Sec.~\ref{sec2} and consider
the form DVs take in the case of the more general spectrum we assume
in this section.

\vskip0.5cm
\subsection{Expansion for large Euclidean momentum}
\lbl{OPE}

Let us begin with a study of the singularity structure of the
Dirichlet series~(\ref{dirichlet}) for $\sigma \to 0^+$.
The expansion around $\sigma=0$ is important because it determines,
through Eq.~(\ref{adler}), the behavior of the OPE for the  Adler
function as  $-q^2\to \infty$. This  includes  the perturbative
series, as the leading term in the OPE.

The mathematical form of the Regge expansion~(\ref{corrections})
is obviously not the most general possible. Therefore, to
ensure that limiting our attention to this form is not overly
restrictive, it is important to show that, at least in principle,
the expansion~(\ref{corrections}) allows us to generate all
the inverse powers and logarithms of $q^2$ present in the OPE.

In order to proceed, it is useful to recall the identity
\be
\lbl{exp}
\mathrm{e}^{-x}=\frac{1}{2 i \pi}\int_C ds \ x^{-s}\,\Gamma(s) \ ,
\ee
where $C$ is a vertical line to the right of $\mathrm{Re}\ s=0$ in
the complex $s$ plane, \ie, to the right of all singularities of
$\Gamma(s)$. One immediately obtains
\be
\lbl{rhomellin}
\borel=\frac{1}{2 i \pi}\int_{\widehat{C}} ds\ \sigma^{-s}
\,\Gamma(s)\, \Phi(s), \quad \mathrm{where}\quad \Phi(s)=
\sum_{n=1}^\infty F(n) \left[  M^2(n)\right]^{-s}\ ,
\ee
where, as a consequence of the asymptotic behavior in
Eq.~(\ref{reggespectrum}), $\Phi(s)$ will have a singularity at
$s=1$, implying that now $\widehat{C}$ is a vertical line to the
right of $\mathrm{Re}\ s=1$.

As it stands, Eq.~(\ref{rhomellin}) is exact. A result, known as the
Converse Mapping Theorem \cite{Flajolet}, relates the behavior of
$\borel$ for $\s\to 0^+$ to the singularities of the function
$\Gamma(s) \Phi(s)$. Since the singularities of $\Gamma(s)$ are
already known to be simple poles located at non-positive integers,
our task is to determine the singularities of $\Phi(s)$.

Let us assume that there is an integer $n^*>0$ large enough such
that the expansion~(\ref{reggespectrum}) applies for $n>n^*$. We
can then split
 \be
 \lbl{splitphi}
 \Phi(s)= \Phi_<(s)+\Phi_>(s)
 =\sum_{n\le n^*}F(n) \left[  M^2(n)\right]^{-s} +
\sum_{n> n^*}^\infty F(n) \left[  M^2(n)\right]^{-s}\ ,
 \ee
and use the expansion~(\ref{reggespectrum}) in the second sum,
$\Phi_>(s)$. Clearly, the function $\Phi_<(s)$ cannot give rise
to any singularity in $s$, hence all singularities are contained
in $\Phi_>(s)$. Let us split
  \be
 \lbl{split}
 \borel=\bl+ \bh
 \ee
where $\bl$ is the inverse Mellin transform of $\Gamma(s)\Phi_<(s)$
and $\bh$ that of $\Gamma(s)\Phi_>(s)$. According to the Converse Mapping
Theorem, since $\Gamma(s)$ contains simple poles at non-positive
integers $s=-k$, $k\ge 0$, the function $\bl$ is of the form
 \be
 \lbl{bl}
 \bl=\sum_{k=0}^\infty \frac{(-1)^k}{k!}\,\Phi_<(-k)\, \s^k\ ,
 \ee
\ie, it is a power series in $\s$. Using Eq.~(\ref{adler}), one sees
that, at least in principle, one obtains all powers of $1/q^2$, as
in the OPE. No logarithms appear yet. All logarithms have to come
from $\bh$ through the singularities of $\Phi_>(s)$, to which we turn next.

Since $\Phi_>(s)$ is defined through a sum over $n> n^*$, we may
insert the expansions in Eq.~(\ref{reggespectrum}) into
Eq.~(\ref{rhomellin}), obtaining
\bea
\lbl{singular}
\Phi_>(s)&=& \sum_{n> n^*} n^{-s}\left(1+ b\, \frac{\log n}{n}
+ \frac{c}{n}+ \frac{\epsilon_M(n)}{n}\right)^{-s}\left(1+
\epsilon_F(n)\right) \\
&= & \Phi_1(s) + \Phi_2(s)\ ,\nn
\eea
where
\bea
\lbl{bigsplit}
\Phi_1(s)& =& \sum_{n>n^*}n^{-s} \big( 1+ \epsilon_F(0,n)\big)
= \sum_{n>n^*}n^{-s}\left(1+\sum_{\nu>0} \frac{d^{(F)}(\nu)}
{\log^{\nu} n}\right)\ ,  \\
\Phi_2(s)&=& \sum_{n>n^*}n^{-s}\left(\epsilon_F(\{\lambda\},n)-
\, \frac{s}{1!} \left(b\frac{\log n}{n}+\frac{c}{n}+
\frac{\epsilon_M(n)}{n}\right) \right.\nn\\
&& \left. \hspace{3cm} +  \frac{s(s+1)}{2!}\left(b\frac{\log n}{n}
+\frac{c}{n}+ \frac{\epsilon_M(n)}{n}\right)^2+ \dots \right) \ . \nn
\eea
The reason for grouping together in $\Phi_1(s)$ the ``1'' with the
terms contained in $\epsilon_F(0,n)$ is because these terms will give
rise to the logarithms of the perturbative series, whereas the
logarithms associated with power corrections will all be contained
in $\Phi_2(s)$.

Let us first deal with $\Phi_1(s)$. One obtains
\bea
\lbl{phione}
\Phi_1(s) &=& \zeta_>(s) + \sum_{\nu>0} d^{F}(\nu) \int_0^\infty dt
\  \frac{t^{\nu-1}}{\Gamma(\nu)}\ \zeta_>(s+t)  \\
&\asymp & \frac{1}{s-1}+  \sum_{\nu>0} d^{F}(\nu) \int_0^\infty dt
\  \frac{t^{\nu-1}}{\Gamma(\nu)}\ \frac{1}{s+t-1}\ ,\nn
\eea
where $\asymp$ means ``singular part of,'' and where
$\zeta_>(s)=\sum_{n>n^*}^\infty n^{-s}$ has the same singularity
structure  as the Riemann $\zeta$-function $\zeta(s)$, whose
singular expansion is $\zeta(s) \asymp \frac{1}{s-1}$. We
substitute Eq.~(\ref{phione})  into Eq.~(\ref{rhomellin}), to obtain the
$\s\to 0^+$ expansion of the $\Phi_1$ part of $\s\bh$.
Interchanging the $s$ and $t$ integrals yields
\be
\lbl{bpt}
\s\bh|_{\Phi_1}= 1+ \sum_{\nu>0} d^{F}(\nu) \int_0^\infty dt
\  \frac{t^{\nu-1}}{\Gamma(\nu)}\ \Gamma(1-t)\,\s^t\
+ \mathcal{O}(\s^{1-\epsilon})\ ,
\ee
with $\epsilon$ a parameter that can be chosen arbitrarily
close to zero.
These manipulations are formal, as the second $t$ integral in
Eq.~(\ref{phione}) diverges at $t\to\infty$ and the $t$ integral in
Eq.~(\ref{bpt}) diverges because of the poles in $\Gamma(1-t)$.
In App.~\ref{dirichletapp} we show that, nonetheless, these
manipulations are valid, specializing to the case $\nu =1$.
It is then straightforward to see that the same argument applies
also for $\nu>1$, and we show this explicitly in App.~\ref{newapp}.

As we will see in Sec.~\ref{sec1prime}, Eq.~(\ref{bpt}) is nothing
but the Borel transform of the usual perturbative series in powers
of $\alpha_s$, in the approximation in which only the first term
in the $\beta$-function is kept. In fact, the poles in $\Gamma(1-t)$
are related to the renormalon singularities associated with the
asymptotic nature of  perturbation theory.  The presence of the
factor $\sigma^t= \re^{t \log\sigma}$ shows that $\borel$ possesses a cut
for $\mathrm{Re}\,\sigma<0$. Using Eq.~(\ref{adler}), this expression
for $\bh$ yields for the perturbative Adler function
\bea
\lbl{adlerpt}
\mathcal{A}(q^2)_{\mathrm{\rm PT}}&\approx& 1+ \sum_{\nu>0}
d^{F}(\nu) \int_0^\infty dt \,\frac{t^{\nu-1}}{\Gamma(\nu)}
\, \frac{\pi t}{\sin(\pi t)}\, \frac{1}{(-q^2)^{t}}\\
&\approx&   1+  \sum_{\nu>0} d^{F}(\nu)  \sum_{k=0}^\infty
\frac{a_k}{\Gamma(\nu)}\ \frac{\Gamma(\nu+k)}
{\left(\log(-q^2)\right)^{\nu+k}}\ ,\nn
\eea
where the identity
\be
\lbl{identity}
\Gamma(1-t) \Gamma(1+t)= \frac{\pi t}{\sin(\pi t)}
\ee
has been used, and the coefficients $a_k$ are defined
by\footnote{An expression of the  coefficients $a_k$ in terms of
the Bernoulli numbers may be obtained, but its precise form is not
very important here.}
\be
\frac{\pi t}{\sin(\pi t)}= \sum_{k=0}^\infty a_k \, t^k\ ,\quad
t\rightarrow 0\ .
\ee

Let us now turn to $\Phi_2(s)$. Its contribution is a linear combination
of terms of the form
\be
\lbl{master}
(-1)^\nu\frac{d^\nu}{ds^\nu}\zeta_>(s+\lambda)=\sum_{n>n^*}
\frac{\log^\nu n}{n^{s+\lambda}}\ ,
\ee
for $\lambda>0$ and we first consider the case $\nu \geq 0$
(we will comment on the case $\nu<0$ below). Inserting this
expression into Eq.~(\ref{rhomellin}) one obtains, after
taking $\lambda$ derivatives with respect to $\s$,
\be
\lbl{master2}
\frac{d^\lambda}{d\s^\lambda}\bh=^{\hspace{-0.23cm}\circ}(-1)^\lambda
\ \frac{1}{2i\pi}\int_{\widehat{C}}\ ds\ \s^{-s-\lambda}
\ \Gamma(s+\lambda)\ (-1)^\nu\frac{d^\nu}{ds^\nu}\zeta_>(s+\lambda)\ ,
\ee
where the symbol $=^{\hspace{-0.23cm}\circ}$ indicates that the
right-hand side is just one of the tems in
$\frac{d^\lambda}{d\s^\lambda}\bh$. Since
\be
(-1)^\nu\frac{d^\nu}{ds^\nu}\zeta_>(s+\lambda)\ \asymp\  (-1)^\nu
\frac{\Gamma(\nu+1)}{ (s+\lambda-1)^{\nu+1}}
\ee
one obtains, after using the residue theorem and integrating
$\lambda$ times over $\s$,
\be
\lbl{masterB}
\bh=^{\hspace{-0.23cm}\circ}\ -\ \frac{(-1)^\lambda}{\nu+1}
\ \frac{\s^{\lambda-1}}{\Gamma(\lambda)}\ (-\log \s)^{\nu+1}
\left(1+\mathcal{O}\left(\frac{1}{\log\s}\right)\right)+
\mathcal{P}_\lambda(\s)\ ,
\ee
where $\mathcal{P}_\lambda(\s)$ is a polynomial of degree $\lambda$.
Substituting this expression into Eq.~(\ref{adler}), the contribution
to the Adler function is given by
\be
\lbl{masteradler}
\mathcal{A}(q^2)=^{\hspace{-0.23cm}\circ} -\ \frac{\lambda}{\nu+1}
\ \frac{\log^{\nu+1}(-q^2)}{(q^2)^\lambda}\  \left(1+\mathcal{O}
\left(\frac{1}{\log(-q^2)}\right)\right)+\mathcal{P}_{\lambda+1}
\left(\frac{1}{q^2}\right)\ .
\ee
As shown in App.~\ref{newapp}, it turns out that the result for
$\nu<0$ may be obtained from (\ref{masterB}, \ref{masteradler}) by
analytic continuation in $\nu$, except for the case $\nu=-1$ which,
due to the singularity at that value, is a special case. For
$\nu=-1$, one obtains (see App.~\ref{newapp})
\be
   \lbl{masterBone}
   \bh=^{\hspace{-0.23cm}\circ}\frac{(-1)^{\lambda-1}}
{\Gamma(\lambda)}\,\s^{\lambda-1}\,\log(-\log\s)\left(1+
   \mathcal{O}\left(\frac{1}{\log(-\log\s)\log\s}\right)\right)
+\mathcal{P}_\lambda(\s)\ ,
   \ee
and  using Eq.~(\ref{adler}),
  \be
  \lbl{masterAone}
   \mathcal{A}(q^2)=^{\hspace{-0.23cm}\circ} -\ \frac{\lambda}{(q^2)^\lambda}
 \ \log\log(-q^2)\left( 1+ \mathcal{O}\left( \frac{1}{\log(-q^2)
\  \log\log(-q^2)}\right) \right)
   +\mathcal{P}_{\lambda+1}\left(\frac{1}{q^2}\right)\ .
\ee

This concludes our exploration of the structure of the OPE generated
by the Regge expansion (\ref{corrections}). Given the variety of
logarithmic corrections obtained, and given the adjustable
parameters $b$, $c$, and $d^{(i)}(\lambda_i,\nu_i)$ in
Eq.~(\ref{corrections}), we conclude that the expansion~(\ref{corrections})
is indeed potentially capable of producing all the necessary terms in
the OPE, including logarithmic corrections, again in the approximation
in which we keep only the leading term in the $\beta$-function. This
is confirmed by the work of Ref.~\cite{Pineda}, which considered this
matching between the spectrum and the OPE in more detail using a more
direct method, and found that the terms in both the perturbative and $1/q^2$
series of the OPE can be matched using the spectrum of
Eq.~(\ref{reggespectrum}). We consider these results good evidence
for the conjecture that, by adjusting the form
of the subleading corrections for $n\to\infty$ in Eqs.~(\ref{reggespectrum})
and~(\ref{corrections}), the complete structure of the OPE, as a
function of Euclidean $q^2$, can indeed be obtained.

\subsection{The perturbative series}
\lbl{sec1prime}
We start with a brief review of the standard Borel summation of the
divergent perturbative series in QCD. Defining the Borel transform
of the perturbative expansion in $\alpha_s$ appearing in
Eq.~(\ref{eq:QCD}) by
\be
\label{eq:BAPT}
B_{\rm PT}^{[{\mathcal A}]}(u)= \sum_{n=0}^\infty  b_n u^n\ ,\qquad  b_n=
\frac{c_{n+1}}{\beta_0^{n+1} \,n!}\ ,
\ee
the perturbative series is formally summed by the Borel-Laplace integral
\be\label{eq:APT}
{\mathcal A}_{\rm PT}(q^2)=1+\int\limits_0^\infty  du\,
\re^{-u /(\beta_0 \alpha_s(-q^2))} \,B_{\rm PT}^{[{\mathcal A}]}(u)\ .
\ee
From renormalon calculus (\ie, the calculation of Feynman diagrams
with bubble insertions) one generically expects
$c_{n+1}\sim \Gamma(n+1)=n!$ \cite{Mueller1992}. Therefore,
because of the $n!$ in the denominator of $b_n$, the
series~(\ref{eq:BAPT}) is expected to be convergent in a
disk $|u|<u_0$ with $u_0>0$ in the Borel plane. If the integral in
Eq.~(\ref{eq:APT}) would be well defined, the original perturbative
series~(\ref{eq:QCD}) would be Borel summable.

Criteria for Borel summability have been formulated in terms of
constraints on the expanded function ${\mathcal A}_{\rm PT}(q^2) $
in the complex $\alpha_s$ plane \cite{Watson}, but these conditions
are not fulfilled in QCD \cite{tHooft}. Borel non-summability is
also manifest because the Borel transform $B_{\rm PT}^{[{\mathcal A}]}(u)$
has singularities along the positive real axis for $u\ge 2$, the
infrared renormalons of Refs.~\cite{tHooft,Mueller1985, Beneke},
which make the integral (\ref{eq:APT}) ambiguous. Other singularities,
the ultraviolet renormalons, are located on the negative real axis
for $u\le -1$, and restrict the convergence of the series (\ref{eq:BAPT})
to the disk $|u|<1$.

It is well-known that, in the large-$\beta_0$ approximation, in which
the coupling is given by Eq.~(\ref{eq:a1loop}), there is a simple
relation between the standard perturbative Borel transform
$B_{\rm PT}^{[{\mathcal A}]}(u)$ of the Adler function and the Borel
transform of the associated spectral function. This can be easily
derived starting from the definition~(\ref{eq:rho}) of $\rho(t)$
and using the Schwarz reflection property, which allows us to write
\be
\rho(t)=\frac{1}{2\pi i}[\Pi(t+i\epsilon)-\Pi(t-i\epsilon)]=
\frac{1}{2\pi i} \int_{C_t}dq^2\, \Pi'(q^2) \ ,
\ee
where $\Pi'(q^2)$ is the derivative of $\Pi$ and  $C_t$ is an open
contour in the complex plane, with end points $t- i\epsilon$ and
$t+ i\epsilon$, which does not cross the cut of $\Pi(q^2)$.
Choosing the contour as a circle of radius $t$ centered at the
origin, parametrized as $q^2=t e^{i\phi}$ for fixed $t$ and
$0\le \phi\le 2\pi$, and using the definition~(\ref{eq:A}), we obtain
\be
\rho(t)=\frac{1}{2\pi}\int_0^{2\pi}d\phi\, {\mathcal A}(t e^{i\phi}) \ .
\ee
Substituting the Borel-Laplace representation~(\ref{eq:APT}) with
the one-loop coupling~(\ref{eq:a1loop}) into this equation and
performing the integral over $\phi$ then yields
\be
\rho_{\rm PT}(t)=1+\int\limits_0^\infty du\, \re^{-u \log (t/\Lambda^2)}
 \,B_{\rm PT}^{[{\mathcal A}]}(u) \, \frac{\sin \pi u}{\pi u}\ .
\ee
Writing a Borel-Laplace representation for the perturbative spectral
function itself,
\be\label{eq:rhoPT}
\rho_{\rm PT}(t)=1+\int\limits_0^\infty du\,\re^{-u/(\beta_0 \alpha_s(t))}
 \,B_{\rm PT}^{[\rho]}(u)\ ,
\ee
we recover the relation
\be\label{eq:BrhoPT}
B_{\rm PT}^{[\rho]}(u) =  B_{\rm PT}^{[{\mathcal A}]}(u) \,
\frac{\sin \pi u}{\pi u}\ ,
\ee
first derived in Ref.~\cite{BrownYaffe}.

For the following discussion, it is useful to establish a relation
between the standard Borel transform introduced in Eq.~(\ref{eq:BrhoPT})
and the new Borel transform $\bh|_{\Phi_1}$ defined in Eq.~(\ref{bpt}).
Indeed, by substituting Eq.~(\ref{eq:rhoPT}) into Eq.~(\ref{rhohat})
and performing the integral over $t$ one finds
\be
\lbl{BPT}
\s\borelPT=1+\int_0^\infty du\,B_{\rm PT}^{[{\rho}]}(u)\
\Gamma(1-u)\ (\sigma \Lambda^2)^u\ ,
\ee
where the subscript PT reminds us of the fact that this relation
holds in perturbation theory. As we will see, the expression
(\ref{BPT}) is nothing but $\s \bh|_{\Phi_1}$  defined in Eq.~(\ref{bpt}).

Our discussion in the previous subsection illustrates clearly how the
OPE corresponds to the branch-point singularity at $\sigma=0$ in the
Borel plane. The purpose of this subsection is to illustrate in more
detail how, in terms of the Borel transform of the perturbative
series reviewed above, the perturbative series appears in our
representation (\ref{adler}) if a general Regge expansion of the
form (\ref{reggespectrum}) is assumed for the spectral function.
We again restrict ourselves to the simplified case in which only the
first coefficient of the $\beta$-function does not vanish.

Let us recall the expansion for the Borel function $\sigma\borel$
as $\sigma \to 0^+$ in Eq.~(\ref{bpt}):
\be
\lbl{pt}
\sigma \bh|_{\Phi_1} = 1 +\sum_{\nu}\dF(\nu)\int_0^\infty \!\!\!du
\,  \frac{u^{\nu-1}}{\Gamma(\nu)}\, \Gamma(1-u)\, \sigma^{u}\ .
\ee
This is the part which contains the perturbative series. To see this,
let us define the combination
\be
\lbl{ptborel}
B^{[\rho]}_{\rm PT}(u)=\sum_{\nu=1}^\infty \frac{\dF(\nu)}
{\Gamma(\nu)}\  u^{\nu-1}\ ,
\ee
so that
\be
\lbl{pt2}
\sigma \bh|_{\Phi_1}  = 1+  \int_0^\infty du\ B^{[\rho]}_{\rm PT}(u)
\  \Gamma(1-u) \, \sigma^{u}\ ,
\ee
where this expression is to be understood as an asymptotic expansion
in $1/\log\sigma$, \ie, as the result of expanding the product
$B^{[\rho]}_{\rm PT}(u)\Gamma(1-u)$ in $u$ about $u=0$ and integrating
term by term.  A comparison of Eq. (\ref{pt2}) with Eq. (\ref{BPT}),
taking into account that we have taken the QCD scale $\Lambda=1$
in (\ref{pt2}), establishes the equality of the two expressions,
as promised above.

The result (\ref{eq:APT}) can be obtained by substituting
Eqs.~(\ref{pt2}) and~(\ref{eq:BrhoPT}) into Eq.~(\ref{adler}),
and, as discussed in the introduction, is valid for
$\frac{\pi}{2}< \arg q^2< \frac{3\pi}{2}$, {\it i.e.},
for $\mathrm{Re}\,q^2<0$, which includes the Euclidean regime.
As we rotate $\sigma$ anticlockwise and reach
$\arg \sigma=\pi-\epsilon$, the integral representation
(\ref{adler}) analytically continues $\mathcal{A}(q^2)$ to
the region $-\frac{\pi}{2}+\epsilon < \arg q^2 < \frac{\pi}{2}+\epsilon$.
Order by order, this continues $\mathcal{A}(q^2)$ through the
perturbative cut at $q^2>0$ into the zeroth Riemann sheet since
the function $\bh$ in Eq.~(\ref{pt2}) is continuous, order by order,
under the corresponding rotation in $\sigma$. The cut discontinuity
of $\bh$ is located further away, at $\arg \sigma =\pi$.

Using the ray $\sigma=|\sigma| \re^{i(\pi-\epsilon)}$, and taking
the limit $\epsilon\to 0$, we can thus define the
perturbative version of the function
$\mathcal{A}(q^2)$ also for $q^2>0$ as
\bea
\lbl{rotateadler}
\mathcal{A}_{\rm PT}(q^2)&=&1+ q^2\int_0^\infty  d|\sigma|\
\re^{-|\sigma| q^2}\int_0^\infty du \ B^{[\mathcal{\rho}]}_{\rm PT}(u)
\ \Gamma(1-u)\ |\sigma|^u \ \re^{i u\pi}\\
&=&1+\int_0^\infty du \  B^{[\mathcal{\rho}]}_{\rm PT}(u)
\ \Gamma(1+u)\Gamma(1-u)\ \frac{\re^{iu\pi}}{(q^2)^u}\nn \\
&=&1+ \int_0^\infty du \  B^{[\mathcal{\mathcal{A}}]}_{\rm PT}(u)
\ \re^{-u\left(\log q^2 -i\pi\right)}\ ,\nn
\eea
which is nothing else than the well-known result for the
analytic continuation to $q^2>0$ of $\log(-q^2)=\log q^2-i\pi$
in the Adler function (\ref{eq:APT}). This exercise illustrates
that the rotation in the complex $\sigma$ plane, supplemented
with the right analyticity properties of the function $\borel$,
produces the right analytic continuation in perturbation theory
of the Adler function  in the $q^2$ complex plane.

We would like to close this section with a comment on the
so-called ``practical version'' of the SVZ sum rules \cite{SVZ},
obtained by neglecting all logarithmic corrections to the
condensates, in the specific case of the vector-channel polarization
considered here. We saw in Eq.~(\ref{adler}) that the function
$\borel$ is given by
\be
\borel= \int_0^{\infty}dt\  \rho(t)\  \mathrm{e}^{-\sigma t}\ ,
\ee
where $\rho(t)$ is the full spectral function defined in
Eq.~(\ref{eq:rho}). There is of course an analogous mathematical
relation between the perturbative counterparts, $\borelPT$ and
$\rho(t)_{\rm PT}$, order by order in powers of
$\alpha_s$.\footnote{The perturbative function $\Pi_{\rm PT}(q^2)$
may exhibit also unphysical singularities, poles or cuts \cite{CaNe},
in the infrared Landau region $-\Lambda_{\rm QCD}^2\leq q^2<0$, which
are not relevant here.} We then see that the practical version of the
SVZ sum rules arises from the assumption that the difference
$\borel-\borelPT$ is an analytic function of $\s$ around the
origin, and  so can be expanded in a power series in $\s$,
$\borel-\borelPT=\sum_{k=0}^\infty c_k\ \s^k $, yielding
\be
\sum_{k=0}^\infty c_k\ \s^k =\int_0^{\infty}dt\ \mathrm{e}^{-\sigma t}
\  \Big(\rho(t)-\rho(t)_{\rm PT}\Big)\  .
\ee
The left hand side is the Borel-Laplace transform of a series in
powers of $1/(-q^2)$ which is also known as the condensate expansion.
In this context, it is traditional to rename the variable $\s\to 1/M^2$
and rewrite the above equation as \cite{SVZ}
\be
\frac{1}{\pi}\int_0^{\infty}dt\ \mathrm{e}^{- t/M^2}
\  \mathrm{Im}\,\Pi(t) =\frac{1}{\pi}\int_0^{\infty}dt
\ \mathrm{e}^{- t/M^2}\ \mathrm{Im}\,\Pi(t)_{\rm PT}
+ \sum_{k=0}^\infty c_k\ \frac{1}{M^{2k}}\ ,
\ee
where $c_k$ is related to the condensate of dimension $2k+2$.
If we assume that the coefficients $d_n$ in Eq.~(\ref{eq:QCD}) are
independent of $q^2$, we obtain from Eq.~(\ref{eq:A}) the
``practical version'' of the OPE,
$\Pi_{\rm OPE}(q^2)=\sum_{n\ge 1} d_n/(n (-q^2)^n)$, from which
the standard SVZ result, $c_k=d_{k+1}/(k+1)!$, immediately follows.

The assumption of analyticity of $\borel-\borelPT$ at the origin
should, however, be treated with some caution. Even though
this practical version of the SVZ sum rules has proved
very successful phenomenologically, and while the logarithmic corrections
to the condensate expansion are screened by at least one power of
$\alpha_s$, these logarithms do exist and render the difference
$\borel-\borelPT$ not analytic in a region around $\sigma=0$.
The phenomenological approximation of neglecting such logarithmic
corrections to the condensate expansion is not guaranteed to work
in general, and should be judged on a case-by-case basis.

\vskip0.4cm
\subsection{Beyond the OPE}
\lbl{nonpert}

We now turn to the singularities of $\sigma \borel$
away from the origin. We saw in Sec.~\ref{sec2} that, when the
asymptotic Regge behavior is exact, {\it i.e.}, $M^2(n)=n$, $F(n)=1$,
the singularities are simple poles located at $\sigma=\pm 2 \pi ik$,
$k=1,2,3,\dots$. What is the fate of these singularities once the
corrections to the spectrum are switched on and the parameters
$b,\, c,\, \epsilon_{F,M}(n)$ in Eq.~(\ref{reggespectrum}) become
non-zero? To study this question, we focus on the region near
the original pole locations. Substituting
$\sigma=\widehat{\sigma}+(\sigma-\widehat{\sigma})$ into
Eq.~(\ref{dirichlet}), implementing the expansions (\ref{reggespectrum})
and (\ref{corrections}), and taking  $\widehat{\sigma}=2\pi i k$, with
$k$ a non-zero integer, we find
\be
\bh=\frac{1}{2 i \pi}\int_{\widehat{C}} ds
\ (\sigma-\widehat{\sigma})^{-s}\ \Gamma(s) \Psi_>(s)
\ee
with
\be
\lbl{psi}
\Psi_>(s)=\re^{-\widehat{\sigma}c}\ \sum_{n>n^*}^\infty
n^{-(s+ \widehat{\sigma}\, b)} \left(1-s\left( b \frac{\log n}{n}
+ \frac{c}{n}\right)+\epsilon_F(n) -\left(\frac{s}{n}
+ \widehat{\sigma}\right)\, \epsilon_M(n)+\dots\right)\ ,
\ee
where we have used that $\re^{-2\pi ikn}=1$. Following steps
exactly analogous to those followed before, we find
\be
\lbl{psising}
\Psi_>(s)\asymp \re^{- \widehat{\sigma}\, c}\,\frac{1}
{s+ \widehat{\sigma}b -1}+\dots\ ,
\ee
which translates into
\be
\lbl{rhohatbranch}
\bh=\re^{-\sh c}\,\frac{\Gamma(1-\sh b)}
{(\sigma-\widehat{\sigma})^{1-\sh b}}+\dots\ .
\ee
The form of this result can be understood by considering the
poly-logarithm representation of the Dirichlet series in the
limit that the $\epsilon_{F,M}(n)$ corrections are neglected.
In that limit, one has
\be
\lbl{polylogdirichlet}
\sum_{n=1}^{\infty} \re^{-\s\left(n+ b\log n+c\right)}=
\re^{-\s c}\sum_{n=1}^{\infty}\frac{\re^{-\s n}}{n^{\s b}}
=\re^{-\s c} \, \mathrm{Li}_{\s b}(\re^{-\s})\ .
\ee
Since
\be
\lbl{polylog}
\mathrm{Li}_{\s b}(\re^{-\s})=\Gamma(1-\s b)
\sum_{k=-\infty}^{k=+\infty}\frac{1}{\left(\s-2\pi ik\right)^{1-\s b}}\ ,
\ee
we see that Eq.~(\ref{rhohatbranch}) corresponds precisely to the
$k$-th term in this sum, reflecting the fact that (\ref{rhohatbranch})
was obtained by expanding $\bh$ in the neighborhood of
$\sigma=\widehat\sigma=2\pi ik$.

Equation~(\ref{rhohatbranch}) shows that the simple pole at
$\sh=2 \pi ik$, present in the exact Regge limit, with
$b=0$, as discussed in Sec.~\ref{sec2}, has now become a branch
point at the same location.

As we show in App.~\ref{app1}, the form of the duality-violating
contribution to the vector-channel polarization now changes from that
given in Eq.~(\ref{adlerminkowskitoy}) to
\be
\lbl{newresult}
\Pi_{\rm DV}(q^2)= 2\pi i\ \sum_{k=1}^{\infty} \re^{i\pi \sh b}
\  (-q^2)^{-\sh b}\ \re^{\sh\left(q^2-c\right)}
+\dots\ ,
\ee
where $\sh=2 \pi ik$ and $q^2>0$. In Eq.~(\ref{newresult}) only the singularities
on the positive imaginary axis contribute, because we rotate
anticlockwise in the $\s$ plane, so that the sum is
restricted to $k>0$. Since
\be
\lbl{prefactor}
(-q^2)^{-\sh b}=(-q^2)^{-2\pi ik b}=
\re^{-2 \pi ikb \log(-q^2-i \varepsilon)}
=\re^{-2k \pi^2 b} \ \re^{- 2 \pi ikb \log|q^2|}\ ,
\ee
we find that
\be
\lbl{Newcorr2}
\Pi_{\rm DV}(q^2)= 2\pi i \sum_{k=1}^{\infty}
\re^{-4 k \pi^2b}\, \re^{ 2  \pi ik \left(q^2- b \log q^2 -c \right)}
\left(1 + \mathcal{O}\left( \frac{1}{\log q^2}\right)\right)\ .
\ee
In Eq.~(\ref{Newcorr2}) we have replaced the dots in
Eqs.~(\ref{psising}) to~(\ref{newresult})
by an explicit estimate of the subleading behavior for large $q^2$.
We see how the change from simple poles to branch cuts, originating
from the logarithmic correction to the spectrum
in Eq.~(\ref{reggespectrum}), leads to logarithmic corrections to the
exponent appearing in the intermediate step of
Eq.~(\ref{adlerminkowskitoy}). The result~(\ref{Newcorr2})
still corresponds to the limit $N_c\to\infty$, but corrections to
a pure Regge spectrum have now been taken into account. If we ignore the
$\mathcal{O}(1/\log{q^2})$ corrections, and set $b=c=0$, we
recover the result of Eq.~(\ref{adlerminkowskitoy}).
We may again enforce the Schwarz reflection property by defining
$\Pi_{\rm DV}(q^2)$ for Im~$q^2<0$ by $\Pi_{\rm DV}((q^2)^*)=
\Pi^*_{\rm DV}(q^2)$.

Rather than discuss this particular result in more detail, we
now proceed to a discussion of how this result gets modified
when $N_c$ is taken large, but finite. We emphasize again the main
message, which is that the simple poles on the imaginary axis in
the Borel plane found in Sec.~\ref{sec2} stay in the same location,
but become branch points instead of simple poles when the spectrum
is generalized to be that of Eq.~(\ref{reggespectrum}).

\begin{boldmath}
\section{DVs for $N_c$ large but finite: A warm-up model}
\lbl{sec3prime}
\end{boldmath}

In Sec.~\ref{sec1} we have seen how the corrections to the asymptotic
Regge spectrum modify the nature, but not the location, of the
singularities of the Borel-Laplace transform $\borel$ in the
$N_c\to \infty$ limit, and how these singularities determine
the form of the DVs. In the following two sections we will
discuss how $1/N_c$ corrections modify these results when
$N_c$ is taken large, but finite.

To this end, it is interesting to first study a physically
motivated model in which all calculations can be carried out
explicitly. The model in question is the one proposed in
Ref.~\cite{Blok}, and it is defined by the correlator
\be
\lbl{shifman}
\Pi(q^2)=\sum_{n=1}^{\infty} \frac{1}{z+n}+ constant\ ,
\ee
where $z=(-q^2)^\zeta$, with $\zeta=1-\mathcal{O}(1/N_c)<1$. This
function has a cut in the  complex $q^2$ plane for $\arg q^2=0$,
and poles on the zeroth Riemann sheet that we may associate with
resonances. Therefore, the model enjoys the analyticity properties
expected in QCD.

In terms of the Borel-Laplace transform, we may write, up to an
infinite real constant,
\be
\lbl{Shifmanborel}
\Pi(q^2)=\int_0^{\infty}d\sigma \ \re^{-\sigma z(q^2)}\ \borel\ ,
\quad \mathrm{Re}\,z>0\ ,
\ee
where $\borel$ is given in Eq. (\ref{rhohattoy}). Since
\be
\lbl{connection}
z=|z|\re^{i \psi}=|q^2| \re^{i\zeta(\varphi-\pi)}\ ,
\quad (0\leq \varphi< 2 \pi \Leftrightarrow \mathrm{1st\ Riemann\ sheet})\ ,
\ee
one finds that a full rotation by an angle $\Delta\varphi=2\pi$ in the
complex $q^2$ plane corresponds to a rotation
$\Delta\psi=\zeta\Delta\varphi=2\pi \zeta< 2\pi$ in the $z$
plane (recall $\zeta<1$), {\it i.e.}, there is a deficit angle.
The poles of the function~(\ref{shifman}) are located at $\psi=-\pi$
in the $z$ plane, which corresponds to $\varphi=\pi - \pi/\zeta <0$
in the $q^2$ plane, \ie, to poles lying on the zeroth Riemann sheet.

Equation~(\ref{Shifmanborel}) is defined for $\arg\sigma=0$ and
$\mathrm{Re}\,z>0$. As we rotate $\sigma$ anticlockwise, going
from $\arg\sigma=0$ to $\arg\sigma=\pi/2-\pi(1-\zeta)=
\pi/2 -O(1/N_c)$,
we can simultaneously rotate $z$ and $q^2$
clockwise, going from $\psi=0$ and $\varphi=\pi$ ({\it i.e.},
Euclidean $q^2$) to $\psi=-\zeta\pi$ and $\varphi=0$ (the Minkowski
regime for $q^2$). At this point, we have not yet reached the poles
of $\borel$ on the imaginary axis in the $\sigma$ plane. Therefore,
this corresponds to a smooth transition in the $q^2$ plane from the
first to the zeroth Riemann sheet, through the cut at $\varphi=0$.

When we keep rotating $\sigma$, at $\arg\sigma=\pi/2$ we encounter
the poles on the imaginary axis of $\borel$. This corresponds to
$\psi=-\pi$ and, through Eq.~(\ref{connection}), to
$\varphi=\pi-\pi/\zeta<0$, {\it i.e.}, to resonance poles
in $q^2$ lying on the zeroth Riemann sheet. It is this correspondence
between the location of the singularities of $\borel$ in the
$\sigma$ plane and the location of the singularities of $\Pi(q^2)$
in the $q^2$ plane that we wish to again emphasize here.
The location of the resonance poles which obstruct the analytic
continuation in $q^2$ and the location of the singularities in $\borel$
which obstruct the analytic continuation in $\sigma$ are linked
through the Borel-Laplace transform.

If we keep rotating $\sigma$ anticlockwise, crossing the
poles on the imaginary axis, we will again pick up the contribution
from the residues of those poles, through Cauchy's theorem,
leading to DVs having the form of the cotangent function in
Eq.~(\ref{adlerminkowskitoy}), with the variable $q^2$ replaced
by $z$.  This is precisely the correct result for DVs
in this model \cite{Blok}.

\vskip0.4cm
\begin{boldmath}
\section{DVs for $N_c$ large but finite: QCD}
\lbl{sec4}
\end{boldmath}
Let us now discuss the effect of $1/N_c$ corrections in the case
of QCD. It is clear that the large-$N_c$ limit must be taken with
care. As we have seen in Eq.~(\ref{Newcorr2}), in the strict
large-$N_c$ limit DVs are not a small correction to the
(analytically continued) OPE. This is not surprising, since
the large-$N_c$ limit of the spectral function is
not a good approximation to the real-world spectral function.

\begin{figure}
\begin{center}
\leftline{\includegraphics*[width=4cm]{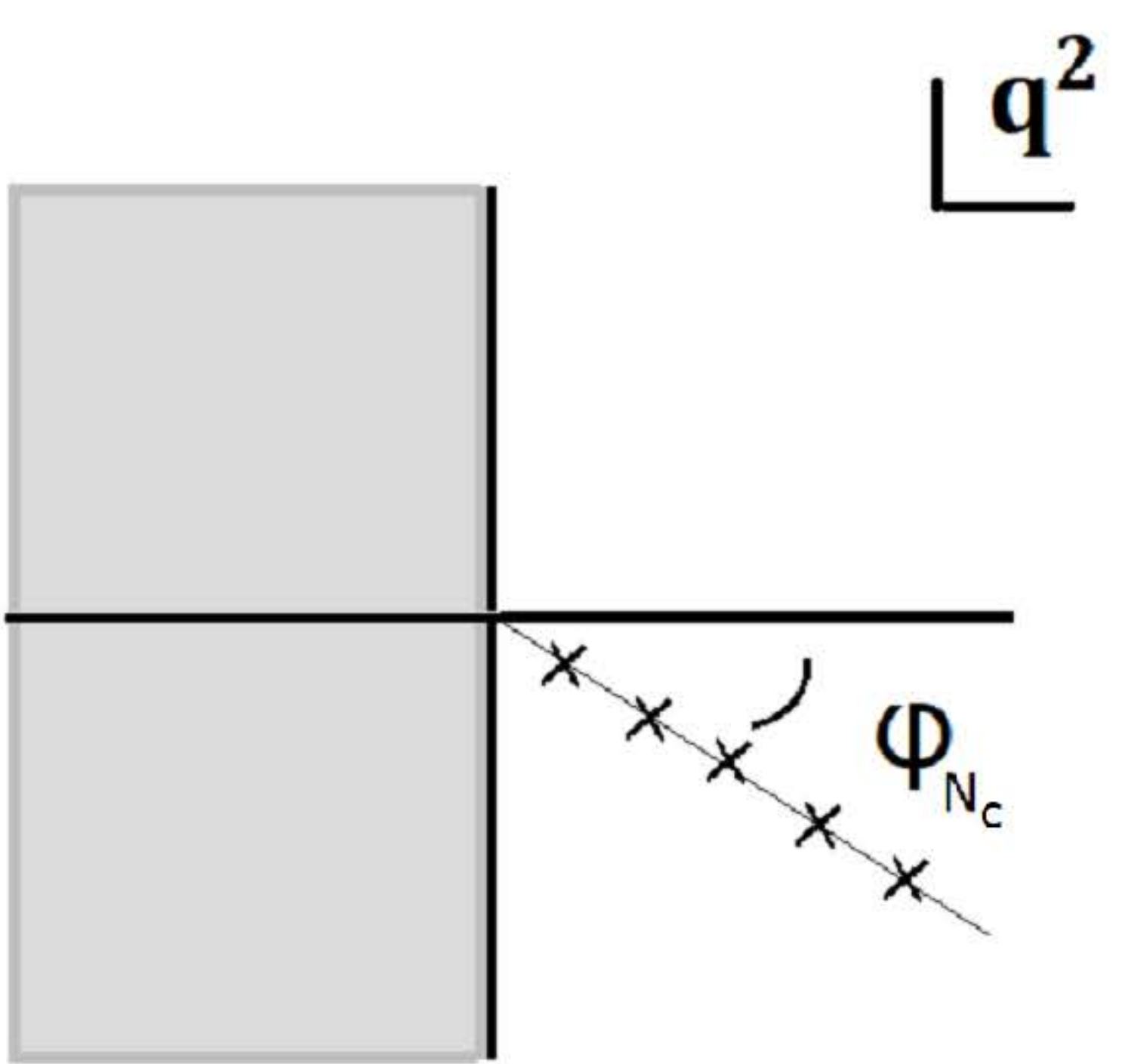}}

\vspace{-1cm}$ \hspace{2cm} \text{ \Huge $\Leftrightarrow$}\hspace{2cm}$

\vspace{-2cm}\rightline{\includegraphics*[width=4cm]{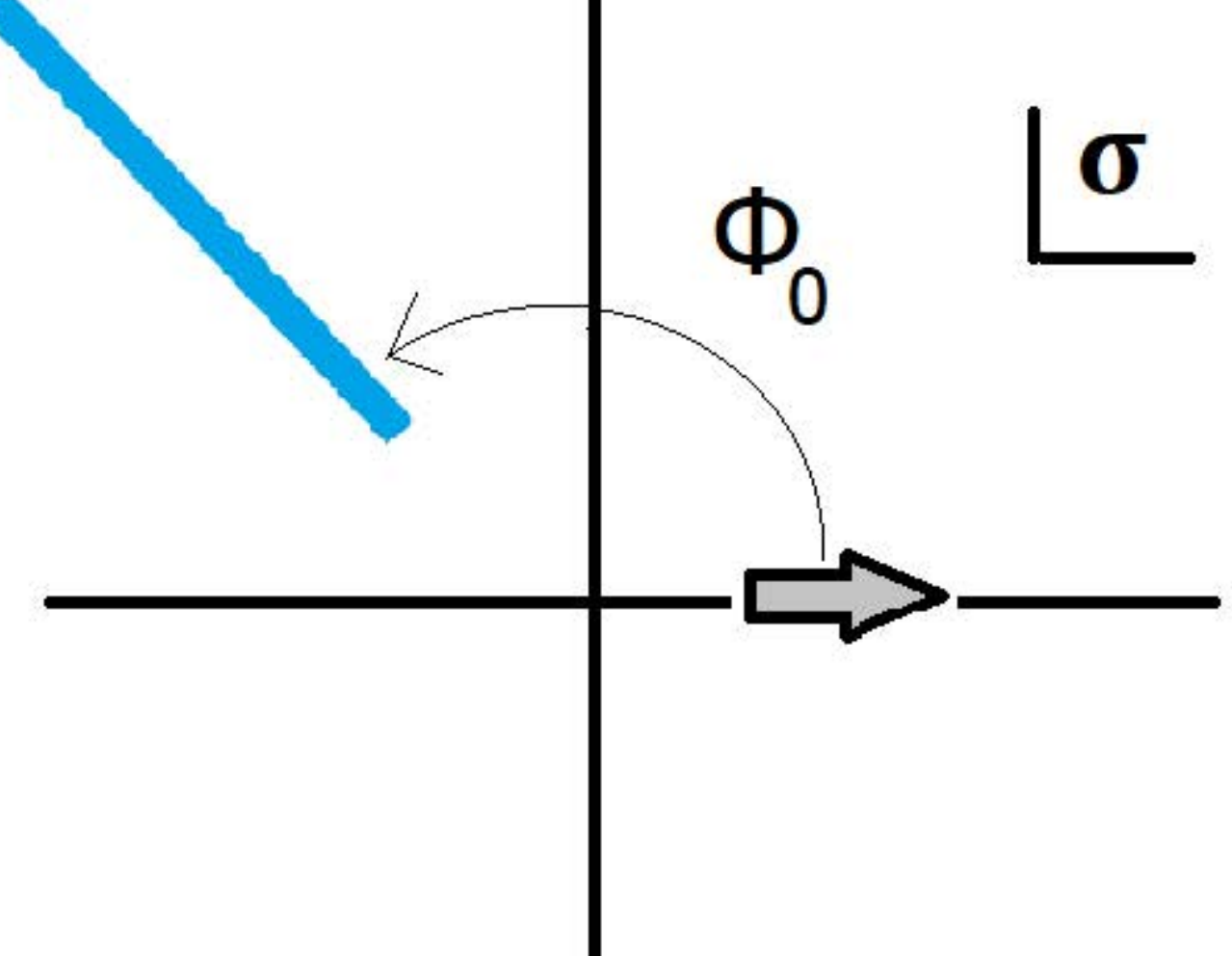}}

\vspace{2.1cm}

\vspace{-1cm}\leftline{\includegraphics*[width=4cm]{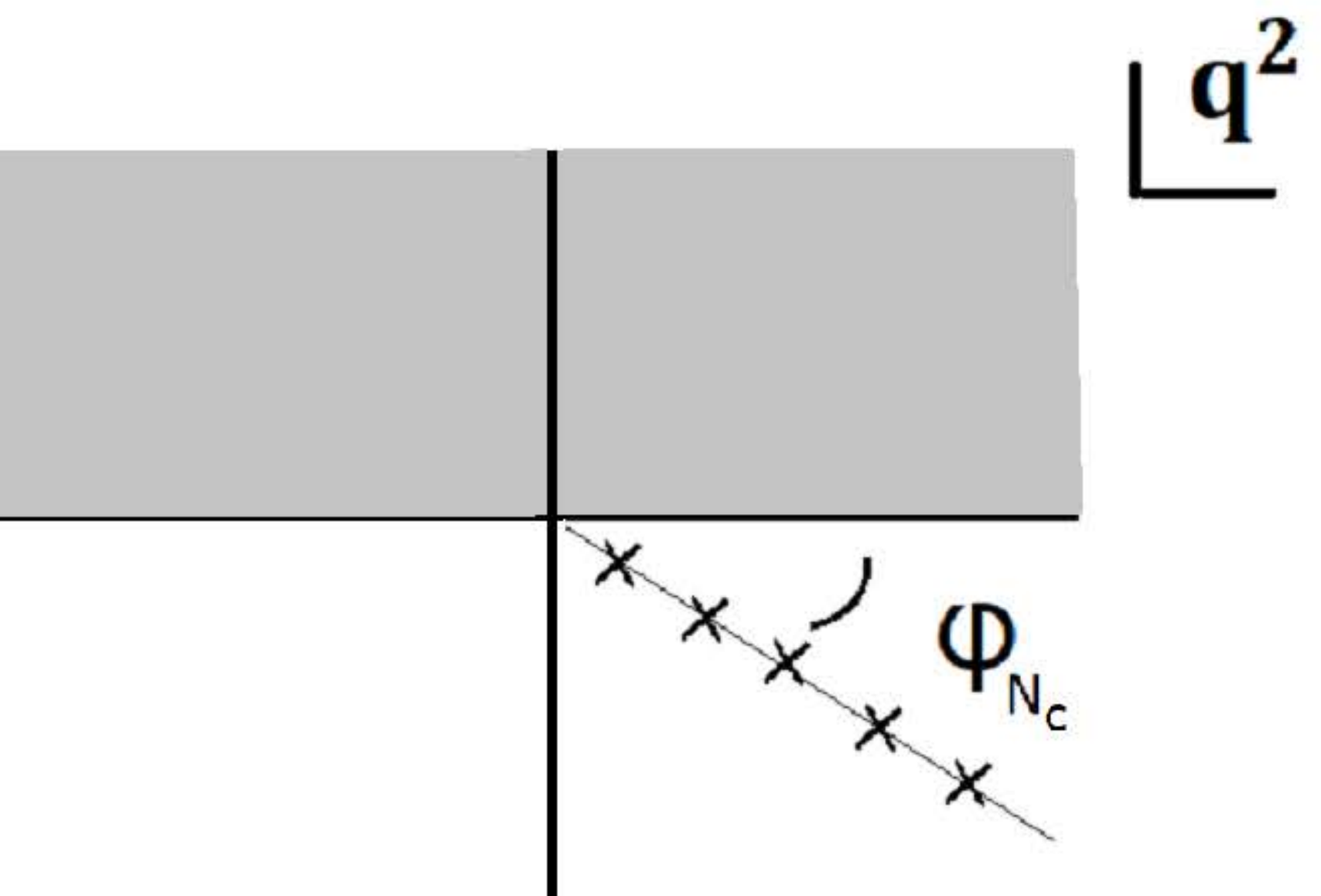}}

\vspace{-1.5cm}$ \hspace{2cm}  \text{ \Huge $\Leftrightarrow$}\hspace{2cm}$

\vspace{-2.2cm}\rightline{\includegraphics*[width=4cm]{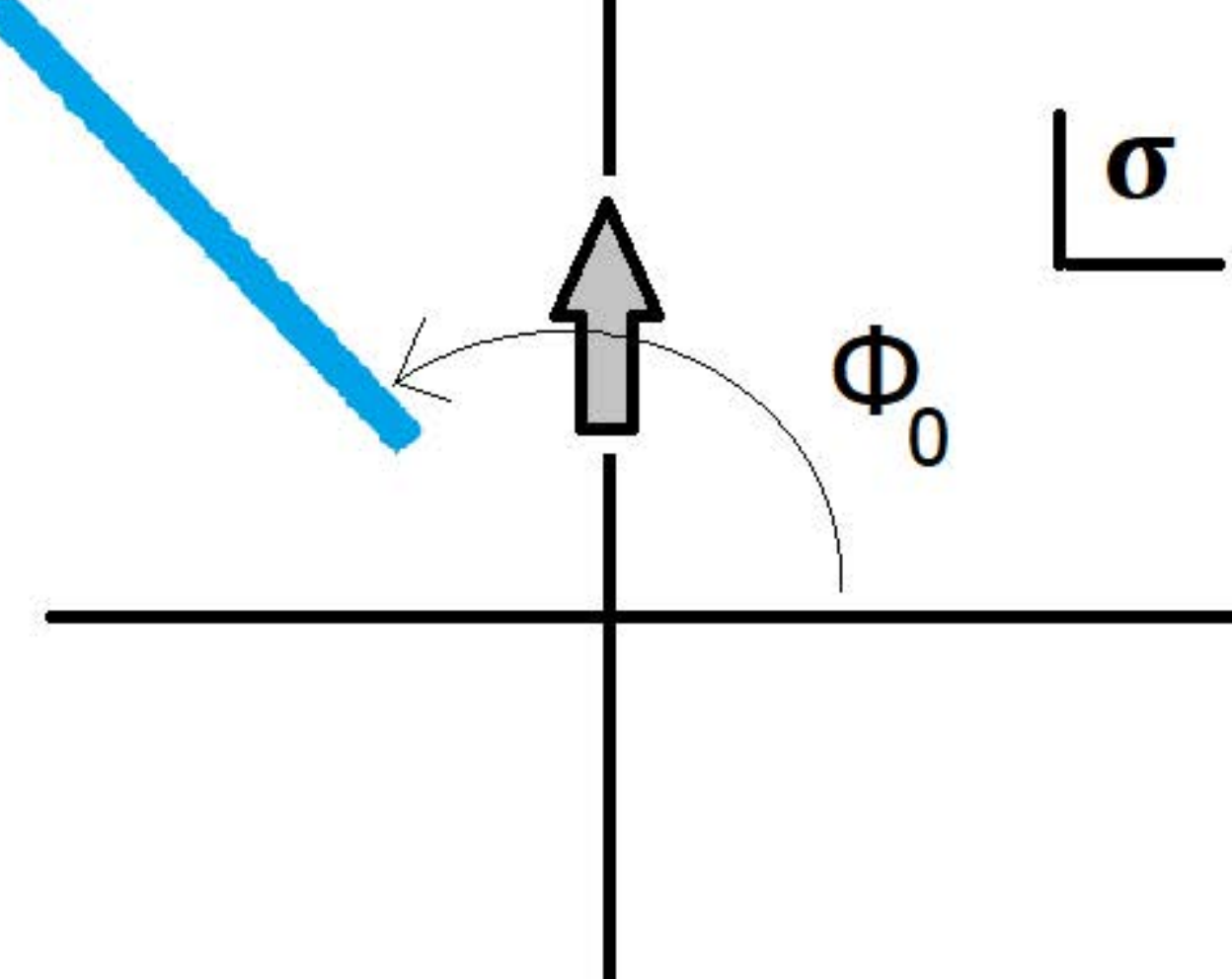}}

\vspace{2cm}

\vspace{-1.5cm}\leftline{\includegraphics*[width=4cm]{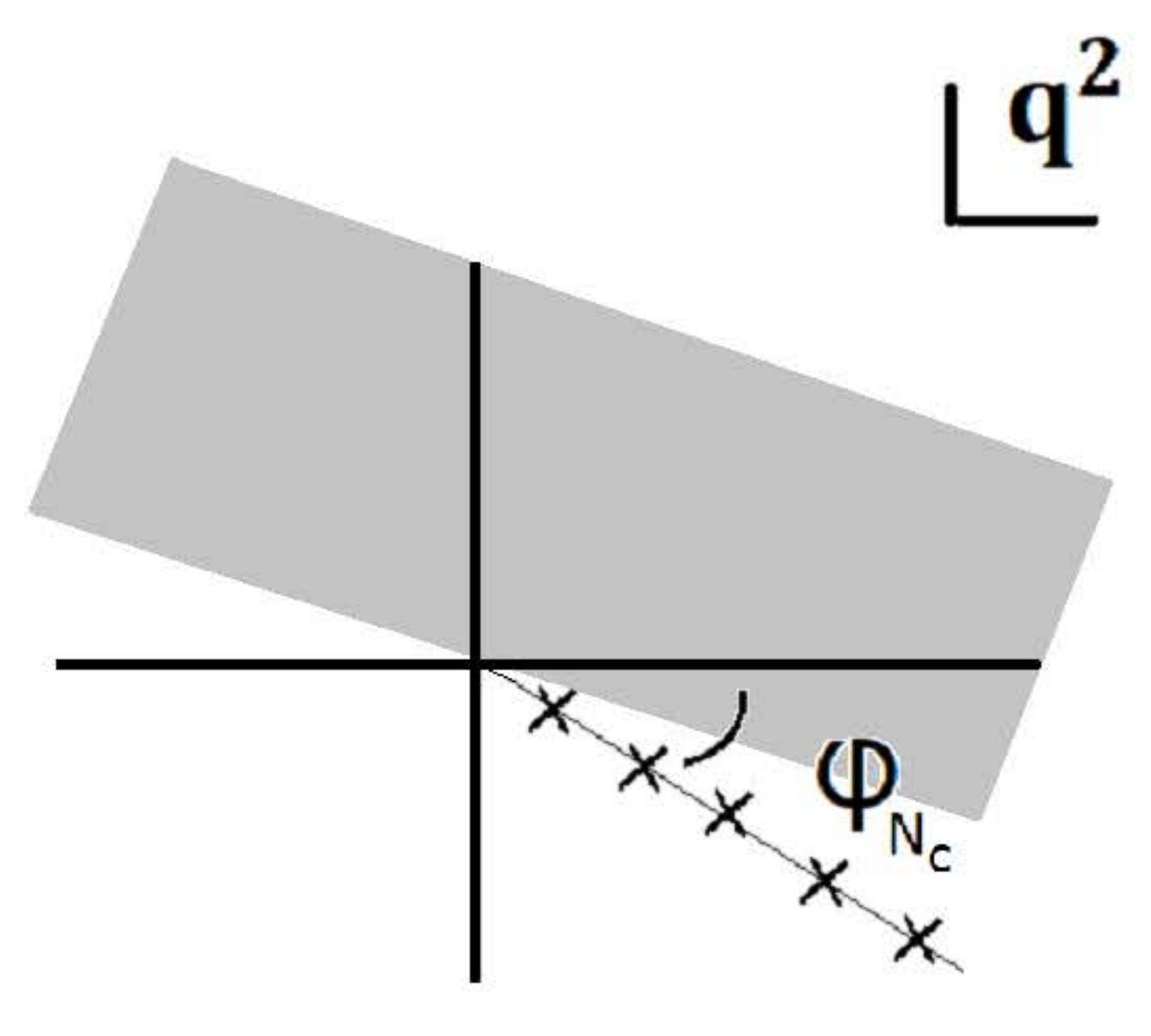}}

\vspace{-1.7cm} $ \hspace{2cm}  \text{ \Huge $\Leftrightarrow$}\hspace{2cm}$

\vspace{-2.cm} \rightline{\includegraphics*[width=4cm]{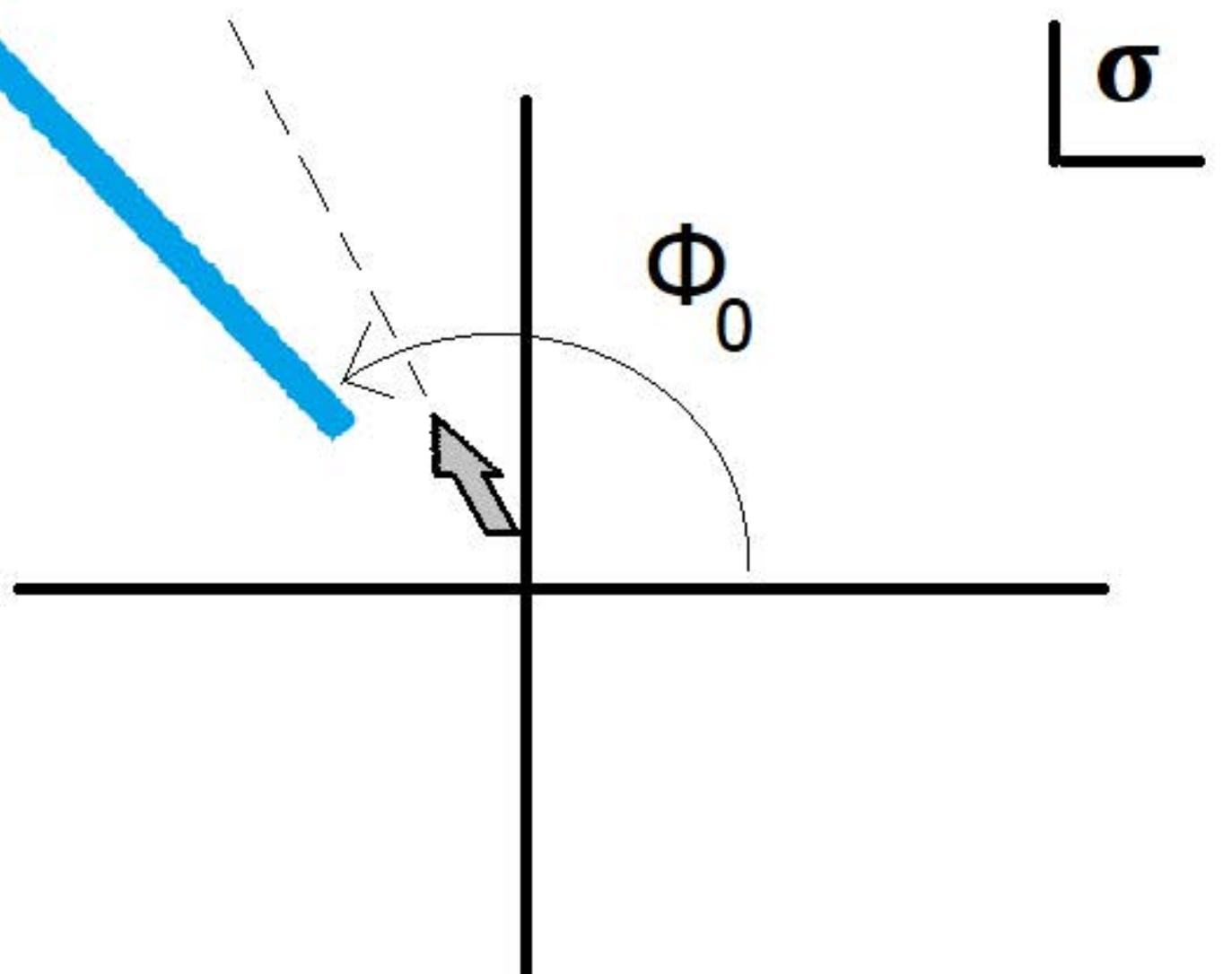}\hspace{-.2cm}}

\vspace{1cm}

\leftline{\includegraphics*[width=4cm]{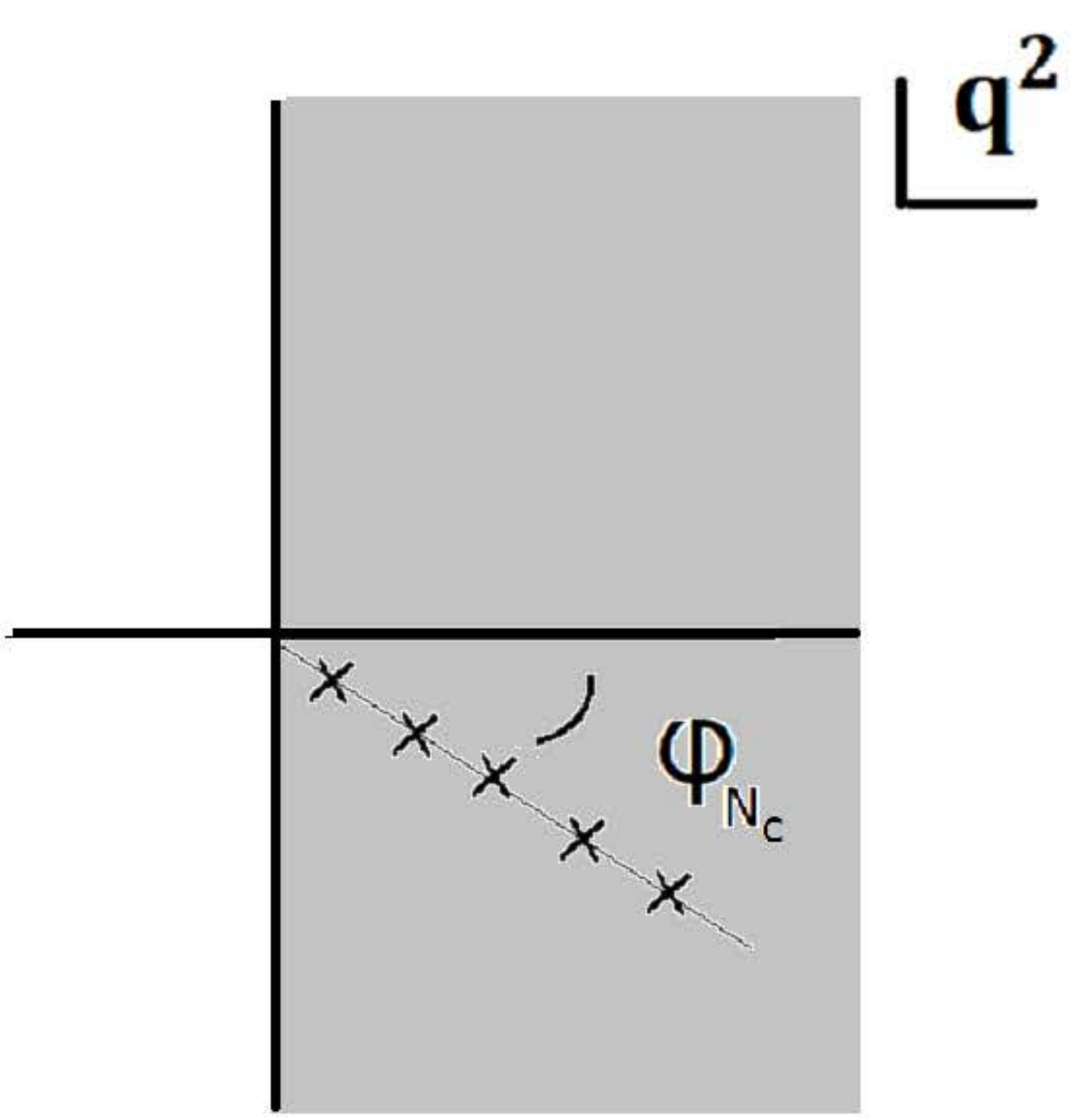}}

\vspace{-2cm}$ \hspace{2cm} \text{ \Huge $\Leftrightarrow$}\hspace{2cm}$

\vspace{-2cm} \rightline{\includegraphics*[width=4cm]{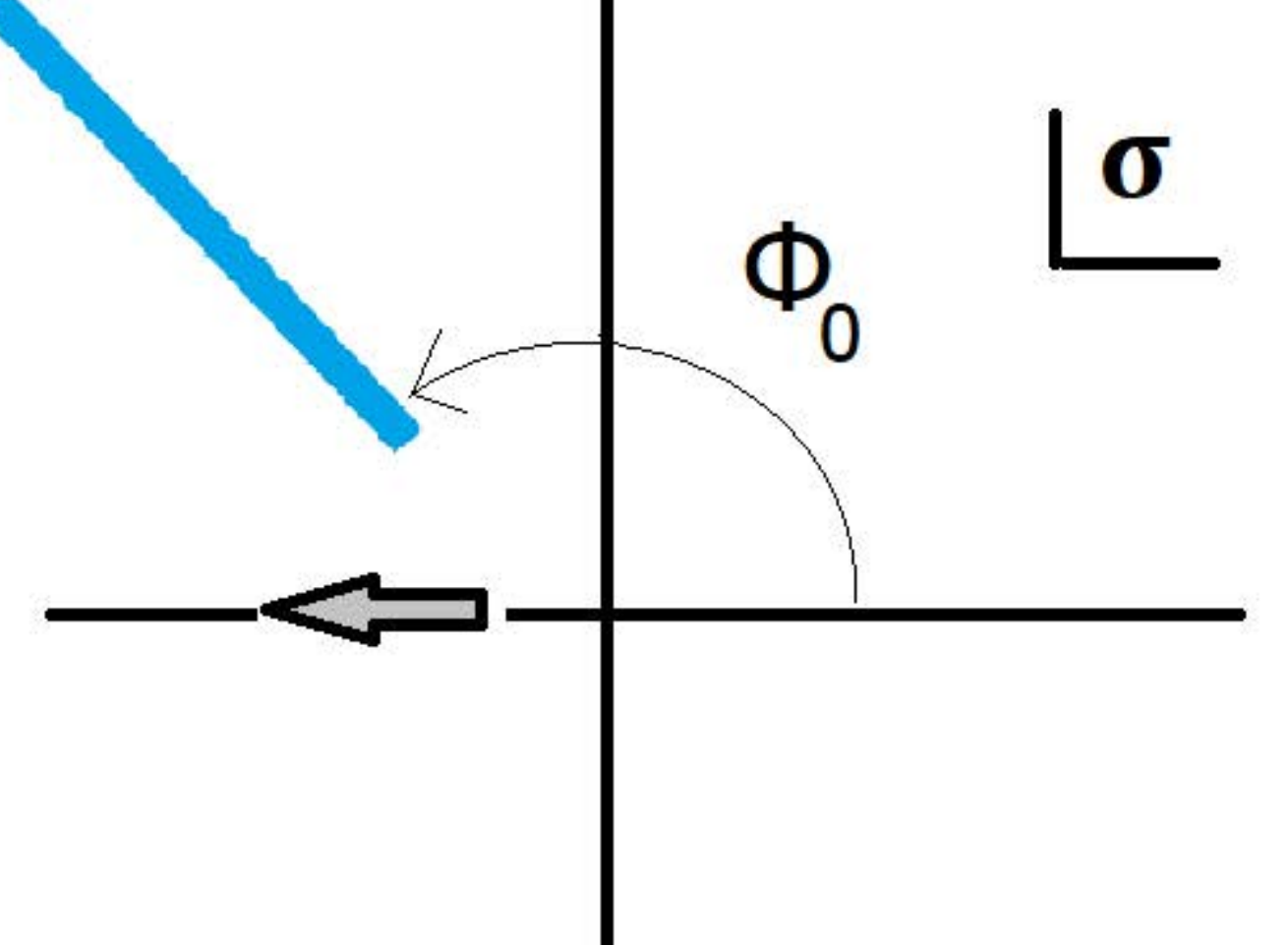}}

\vspace{0.8cm}

\includegraphics*[width=5cm]{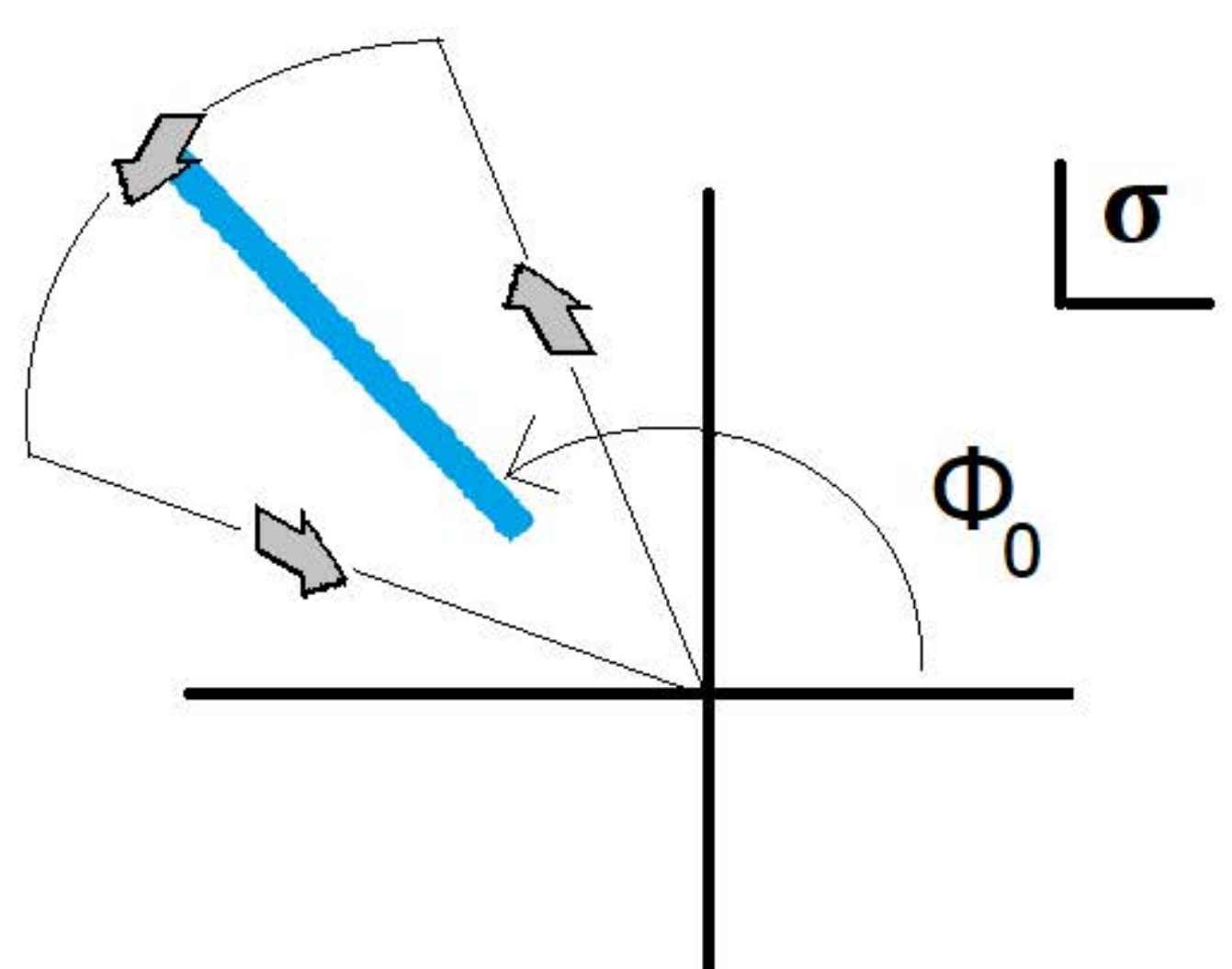}

\caption{Contour used to jump over the cut in the $\sigma$ plane.
Analytic extension into the zeroth Riemann sheet using the generalized
Borel-Laplace transform.}
\end{center}
\end{figure}

The spectrum of QCD is not known in any detail at large, but finite,
$N_c$, so we can no longer calculate the function $\sigma\borel$
from Eq.~(\ref{rhohat}) as we did in the previous sections. Some
important qualitative features of the spectral function are
known, however. Moving away from the strict large-$N_c$ limit to $N_c$ large
but finite, it is known, for example, that the poles of $\Pi(q^2)$
on the Minkowski axis move a small distance away into the zeroth
Riemann sheet and a cut in $\mathrm{Im}\,\Pi(q^2)$ appears on this axis.

Starting from the initial representation~(\ref{adler}) with
$\arg \sigma=0$, valid for $\frac{\pi}{2}< \arg q^2< \frac{3\pi}{2}$,
as we rotate towards $\arg \sigma= \frac{\pi}{2}+ \epsilon$ and cover
the region $-\epsilon< \arg q^2 < \pi - \epsilon$, now nothing dramatic
occurs. In contrast to the case of the strict large-$N_c$ limit,
where the resonance poles are located on the Minkowski axis, now that
$N_c$ is finite, as we move from $\arg q^2= + \epsilon$ to
$\arg q^2= -\epsilon$, crossing the Minkowski axis, we move into
the zeroth Riemann sheet without encountering any singularity.
This is so because $\Pi(q^2)$, as we saw in the warm-up model model
in Sec.~\ref{sec3prime}, and in the perturbative series in
Sec.~\ref{sec1prime}, is \emph{continuous} across the
corresponding cut. It is only as $\arg q^2$ becomes more
negative, and $q^2$ moves deeper into the zeroth Riemann
sheet, that the poles corresponding to the presence of
resonances are encountered. When $N_c$ is large (but finite)
a resonance pole in the complex plane is located at
an angle $\varphi_{N_c}$ given by
\be
\lbl{angle}
\tan \varphi_{N_c}\approx \varphi_{N_c}  =- \frac{\Gamma}{M}= -
\frac{a}{N_c}\left(1 + \mathcal{O}\left(\frac{1}{N_c}\right)\right)\ ,
\ee
where $a\sim N_c^0>0$ and we have used that $\Gamma\sim 1/N_c$ and
$M\sim N_c^0$. A string-based model suggests that the parameter
$a$ is independent of the resonance excitation number
$n$ \cite{string}.\footnote{This feature is also observed
in two-dimensional QCD \cite{Blok}.} Thus, as $1/N_c$ corrections
cause the resonance poles to rotate clockwise by an angle
$\varphi_{N_c}\approx - \frac{a}{N_c}$ in the complex $q^2$ plane,
the singularities of $\borel$ in the complex $\sigma$ plane,
according to Eq.~(\ref{adler}), rotate anticlockwise by the same angle
past the positive imaginary axis. These singularities must therefore be
located (approximately) along the ray
$\arg \sigma \simeq  \frac{\pi}{2}+ \frac{a}{N_c}\equiv\phi_0$
(see Fig.~2), where we have assumed here that $a$ does indeed not
depend on $n$. In fact, we are primarily interested in the
singularity closest to the origin, as this will be the one
that generates the leading contribution to the DVs. A mild dependence
of $a$ on $n$ will thus have no impact on our conclusions.

If the large-$N_c$ limit is a reasonably smooth one, the
distance of this closest singularity to the origin cannot be too
different from that found in the $N_c=\infty$ limit in
Sec.~\ref{sec1}, namely, $|\sh(k=1)|=| 2\pi i|=2\pi$. We thus assume
that the singularity closest to the origin is located at
$\sh= \sigma_0 \ \mathrm{e}^{i \phi_0}$ with
$\phi_0=\frac{a}{N_c}+ \frac{\pi}{2}$ and $\sigma_0=2\pi$,
up to subleading $1/N_c$ corrections, as depicted in Fig. 2.

In this case, as $\sigma$ is rotated from $\arg \sigma=0$ to
$\arg \sigma=\pi$, we cross the branch cut (depicted by a
blue line in Fig. 2). This generates a DV contribution given by
the integral over the contour, $\Gamma$, shown in the bottom panel
of Fig. 2, akin to that appearing in Eq.~(\ref{adlerDV}) of
Sec.~\ref{sec2}, and having the form
\be
\lbl{DV}
\frac{d\Pi_{\rm DV}}{dq^2}(q^2)= \int_{\Gamma}d\sigma\
\mathrm{e}^{\sigma q^2} \sigma \borel\ ,
 \ee
where we can now take $q^2>0$.  Equivalently,
 \be
 \lbl{DVpi}
\Pi_{\rm DV}(q^2)=\int_{\Gamma}d\sigma\
\mathrm{e}^{\sigma q^2}\borel\ ,
 \ee
up to a constant of integration which, as in Sec.~\ref{sec2},
has no physical effect.

Let us assume that the function $\borel$ is of the general form
 \be
 \lbl{branch}
 \borel=\frac{a_0}{(\sigma-\sh)^{1+\gamma}}\left[ 1+
a_1 \left (\s-\sh\right)^{p_1}+ \frac{a_{\log}}{\log(\s-\sh)}+\dots\right]\ ,
 \ee
where $\gamma=-\sh b + \mathcal{O}(N_c^{-1})$, and
$a_0=\re^{-\sh c}\ \Gamma(1+\gamma)\left(1+
\mathcal{O}\left(N_c^{-1}\right)\right)$, in accordance with
Eq.~(\ref{rhohatbranch}) in Sec.~\ref{nonpert}. The generic parameters
$p_1>0$ and $a_1, \, a_{\log}$ encapsulate
the dependence on the corrections to the Regge spectrum associated
with the quantities $\epsilon_{F,M}$ in Eq. (\ref{psi}).
One then obtains for the associated duality-violating contribution
 \be
 \lbl{DVbranch}
  \Pi_{\rm DV}(q^2)=\int_{\sigma_0\ \mathrm{e}^{i \phi_0}}^{\infty\
\mathrm{e}^{i \phi_0}}d\sigma \ \mathrm{e}^{\sigma q^2}
\mathrm{Disc}\{  \borel\} \ ,
\ee
which yields ({\it cf.} App.~\ref{app1})
\be
\lbl{result}
\Pi_{\rm DV}(q^2)= 2  \pi i\ \re^{-2\pi i\left(c+\gamma/2\right)}
\, (-q^2)^\gamma \ \mathrm{e}^{q^2\sh}\left[1+ a_1 \,
\frac{\Gamma(1+\gamma)}{\Gamma(1+\gamma-p_1)}\,
\frac{\re^{i \pi p_1}}{(-q^2)^{p_1}} + \frac{a_{\log}}
{\log q^2}+\dots  \right]\ .
\ee
This result depends solely on the location of the branch point
($\sh=\sigma_0 \ \mathrm{e}^{i \phi_0}$) and the nature of the
branch cut, $\gamma$. The orientation of the branch cut in the complex
plane is irrelevant, as expected.

For $N_c$ large, recalling that $\sh=\sigma_0\left( - \sin(\frac{a}{N_c})
+ i \cos(\frac{a}{N_c})\right)= 2\pi (i - \frac{a}{N_c})
\left(1+\mathcal{O}(N_c^{-1})\right)$ and
$\gamma=-2\pi ib+ \mathcal{O}(N_c^{-1}) $, one obtains
\bea
\lbl{resultNc}
\Pi_{\rm DV}(q^2)&=&  2 \pi  i \, \re^{-2\pi^2 b}\, \re^{-2\pi ci}\,
(-q^2)^{-2\pi ib}\ \re^{ 2\pi q^2 \left(i-\frac{a}{N_c}\right)}\ \nn\\
&&\hspace{-1cm}\times  \left[1+ a_1 \, \frac{\Gamma(1-  2\pi ib)}
{\Gamma(1-2\pi i b-p_1)}\, \frac{\re^{i \pi p_1}}{(-q^2)^{p_1}}
+ \frac{a_{\log}}{\log q^2} +\dots   \right]+
\mathcal{O}\left(\frac{1}{N_c}\right)\ ,
\eea
from which one can extract the leading contribution
\bea
\lbl{final result}
 \hspace{-3cm}\Pi_{\rm DV}(q^2)&\approx & \ 2\pi i \,  \re^{-4\pi^2 b}\,
\re^{-2 \pi q^2 \frac{a}{N_c}} \\
  &&  \times \bigg[\cos2\pi\Big(q^2- c- b \log{q^2}\Big)
+ i \sin 2\pi \Big(q^2-c - b\log{q^2}\Big)\bigg]\   \nn\\
   &&\hspace{2cm}\left(1+ \mathcal{O}\left(\frac{1}{N_c};
\frac{1}{(q^2)^{p_1}};\frac{1}{\log q^2}\right)\right)\nn\ ,
 \eea
where Eq.~(\ref{prefactor}) has again been used. Since
Eq.~(\ref{final result}) is valid for $q^2>0$, taking the
imaginary part yields the DV part of the spectral function. As
before, Eq.~(\ref{resultNc}) also gives $\Pi_{\rm DV}(q^2)$ in
the complex plane, for Re~$q^2>0$ and Im~$q^2>0$. For Im~$q^2<0$
it is defined using $\Pi_{\rm DV}((q^2)^*)=[\Pi_{\rm DV}(q^2)]^*$,
thus enforcing the reflection property. The new contribution
$-q^2 d\Pi_{\rm DV}(q^2)/dq^2$ should be added to Eq.~(\ref{eq:QCD})
in order to obtain a complete representation of the Adler function
in the Minkowski region, for large $q^2$.

As one can see, the main effect of the subleading terms in the
Regge expansion at large $n$ is the logarithmic correction
to the argument of the cosine and sine functions modulating the exponential
falloff with $q^2$. There are at least two reasons to expect these corrections to
generate only small modifications to these sinusoidal factors. First,
$|b\log q^2|\ll q^2$ for any $b$ at large $q^2$. Second,
the phenomenological knowledge which, as we have said,
supports a Regge behavior in QCD, does not yield any evidence
for a non-zero value for the $\log n$ term in the mass spectrum.
In other words, phenomenology is consistent with a small $b$ in
QCD. This result provides theoretical support for the
parametrization introduced in \cite{Cata1,Cata2}, which
was successfully tested against precise data for the non-strange
vector and axial-vector spectral functions obtained from
hadronic $\tau$ decays by the OPAL \cite{OPAL} and
ALEPH \cite{ALEPH} experiments, in a series of analyses of
the QCD coupling, $\alpha_s$ \cite{Boito1,Boito2,Boito3}.
In App.~\ref{sec:numerics} we give some numerical evidence
for the agreement between the results obtained from the fits
to Regge trajectories obtained, {\it e.g.}, in Ref.~\cite{Masjuan},
and those obtained from fits to the $\tau$ data.
Further theoretical studies using the functional analysis methods
developed in \cite{CGP14, BC2017} may also help understanding the
origin and nature of DVs in QCD.

We end this section with two comments. First, as already noted
at the end of Sec.~\ref{sec2}, even for Euclidean $q^2$, the OPE
for $\Pi(q^2)$ is asymptotic, and, at any finite order the remainder
is of order $\mbox{exp}(-R|q^2|)$, where $R$ is the distance of the
nearest non-perturbative singularity in the Borel plane. Here, up
to $1/N_c$ corrections, $R=2\pi$, and thus these exponential remainders
are much smaller than the exponential suppression factor
$\mbox{exp}(-2\pi q^2 a/N_c)$ in Eq.~(\ref{final result})
for $N_c$ large enough. This singularity thus plays two roles:
the absolute value of its position in the complex plane
sets the size of the exponential remainder for the OPE in the Euclidean
regime, while both the absolute value and its phase determine
the form of the DV contributions in the Minkowski regime. This
can be explicitly verified in the simple model of Sec.~\ref{sec2},
for example. In the more realistic approach of the subsequent
sections, also a logarithmic cut starting at $\sigma=0$ in the
Borel plane appears. However, this cut plays a different role, leading
to the logarithmic corrections in each term in the OPE,
as discussed in Sec.~\ref{OPE} above.
Second, if we
take $q^2=s+i\Delta$ in Eq.~(\ref{resultNc}),
we find that DVs are exponentially suppressed with a factor
$\mbox{exp}[-2\pi\Delta]$ away from the positive real $q^2$
axis.   If we then follow the prescription of Ref.~\cite{PQW} by
taking $\Delta\propto s$, DVs are exponentially suppressed
at large $s$, even in the limit $N_c\to\infty$.
Our results are thus consistent
with the smearing method proposed in Ref.~\cite{PQW}.

\vskip0.4cm
\begin{boldmath}
\section{Conclusions}
\lbl{sec:conclusions}
\end{boldmath}
In this paper, we analyzed the large $q^2$ behavior of the Adler
function, with the aim of deriving its form on the Minkowski axis,
where neither perturbation theory nor its supplemented version,
represented by the full OPE, provide a reliable representation.
While the OPE is the dominant contribution at
large $q^2$, there are additional nonperturbative contributions,
which are not part of the OPE. These quark-hadron duality violating
contributions can be probed starting from fairly general assumptions
about the spectrum using the techniques of complex analysis. Our
main result is the expression for the leading duality-violating
contribution to the vacuum polarization, given in Eq.~(\ref{final result}).

For any analysis such as ours, some non-perturbative input
going beyond the OPE is needed. This non-perturbative input should
reflect the properties of the spectrum, which determines the Adler
function through the dispersion relation, Eq.~(\ref{adler}). This
relation also shows that the Adler function is a function of one
variable, $q^2$ (as long as we work in the chiral limit), expressed
in terms of the scale of QCD. However, in practice, by introducing
the perturbative coupling $\alpha_s(-q^2)$, it is usually rearranged
in terms of a double expansion in powers of $\alpha_s(-q^2)$ and $1/q^2$,
given a choice of renormalization scheme.

Our analysis is based on the fact that we can write the Adler function
as the Borel transform, in the
plane of the complex variable $\sigma$, of a function
$\sigma B^{[\rho ]}(\sigma )$, where $B^{[\rho ]}(\sigma )$ is, itself, the
Laplace transform of the spectral function, {\it cf.} Eqs.~(\ref{adler})
and~(\ref{rhohat}). The Borel-Laplace transform $\borel$ allows us to
effectively continue the asymptotic expansion of the OPE from the
Euclidean to the Minkowski domain. This is accomplished by
taking advantage of the combination of the exact nature of the
dispersion relation~(\ref{adler}) and the powerful
techniques of analytic continuation. The Borel-Laplace representation,
moreover, allows us to relate the large Euclidean $q^2$ behavior of
the Adler function to the behavior of $\borel$ near $\sigma=0$. In
particular, the logarithmic corrections to the OPE are directly
related to the cut along the negative real axis
emanating from $\sigma=0$ in the Borel plane, as shown in
Secs.~\ref{OPE} and~\ref{sec1prime}. In Sec.~\ref{sec1prime} we
also recovered the standard renormalon picture relating the OPE
to perturbation theory, and rederived the SVZ sum rules.

There can be no singularities to the right of the imaginary axis
in the $\sigma$ plane, as follows directly from Eq.~(\ref{adler}).
However, we find that, beyond the singularity at $\sigma=0$, there
may be further singularities in the half-plane Re~$\sigma\le 0$,
with the location and nature of these singularities depending on
general properties of the spectrum.

Since the full spectrum of QCD\footnote{Here, we are of course
concerned with the channel relevant to the vector-current only.}
is not known, even in the large-$N_c$ and
chiral limits, we have had to make assumptions in order to be able to
identify the location and nature of these singularities.  Our main
assumption is that, for asymptotically large energies, the spectrum
for $N_c=\infty$ lies on a Regge trajectory.\footnote{Technically,
a radial trajectory.} This assumption is supported by phenomenology,
intuitive arguments based on string theory and the solution of
large-$N_c$ QCD in two dimensions. At large but finite energies,
we parametrize the spectrum in terms of a rather general form,
with many parameters ($b$, $c$ and the parameters $d^{(F,M)}(\nu_{F,M})$
and $d^{(F,M)}(\lambda_{F,M},\nu_{F,M})$ of Eqs.~(\ref{reggespectrum})
and~(\ref{corrections})). Starting from the limit $N_c\to\infty$,
it turns out to be possible to extend the analysis to large but
finite $N_c$, with plausible additional assumptions
(see Sec.~\ref{sec4}).

While we cannot derive these assumptions from QCD, we
{\it can} show that, starting from these assumptions, it is
possible to reconstitute the OPE for large Euclidean $q^2$. Explicitly,
with the general parametrization of the spectrum given by
Eqs.~(\ref{reggespectrum}) and (\ref{corrections}), there is enough
freedom available to allow a match to the usual form of the OPE, where
inverse logarithms can be re-expressed in terms of $\alpha_s(-q^2)$ in
the large-$\beta_0$ approximation. This result can be generalized
to include also higher-order terms in the $\beta$ function affecting
the relation between $q^2$ and $\alpha_s(-q^2)$. In fact, our results
agree with those of Ref.~\cite{Pineda}, where also some contributions
beyond the large-$\beta_0$ approximation were considered.
While we have not traced the contribution
of all such needed additional Regge spectrum corrections to our
final result, Eq.~(\ref{final result}), we conjecture that such
corrections will not alter the shape of the leading expression for
$\Pi_{\rm DV}(q^2)$.

We find that for $N_c\to\infty$, the singularities of $\borel$ are located
on the imaginary axis, while for finite $N_c$ they rotate anticlockwise
from the imaginary $\sigma$ axis by an amount $\sim 1/N_c$, associated
with the decay widths of resonances. The singularity closest to the
origin, at a distance approximately equal to $2\pi$ in units of the
Regge slope, yields the leading term in the duality-violating
contribution to the vacuum polarization, Eq.~(\ref{final result}).
Singularities farther away lead to exponentially subleading terms.
These singularities are unlike the cut starting at $\sigma=0$ along
the negative real axis, which is associated with the perturbative
expansion (and perturbative corrections to the higher-order terms in
the OPE), as discussed in Sec.~\ref{sec1prime}. In this sense, the
two types of singularities in the Borel plane, and, therefore, the
two expansions, are clearly separated.

We conjecture that the existence of these singularities in the Borel
plane is more general than just a consequence of the Regge behavior,
with corrections of the form assumed in this paper. These
singularities in the $\sigma$ plane are a direct consequence of the
fact that the spectral function extends all the way to infinity: if
the spectral function were to vanish beyond a finite value,
$t_{max}$, of $t$, there would be no singularities in $\sigma$, and
the OPE would be a convergent power series in $1/q^2$ (for $q^2>t_{max}$)
without any corrections logarithmic in $q^2$. Thus, the fact that
the OPE is divergent suggests that there are contributions which
are exponentially suppressed in the inverse of the expansion variable,
$1/q^2$, in accordance with the notion of a transseries, and this
is precisely what we find to be the case.

There are several questions we have not answered. One obvious question is
whether our analysis can be extended systematically beyond the class
of corrections to a Regge spectrum given by Eqs.~(\ref{reggespectrum})
and~(\ref{corrections}), and, connected to that, beyond the
large-$\beta_0$ approximation. Another physically interesting question
is how our results would change when a non-vanishing quark mass is taken
into account. These questions are beyond the scope
of the current paper.

\vskip0.4cm
\section*{Acknowledgements}

MG and SP would like to thank M. Jamin, P. Masjuan and A. Pineda
for conversations.  SP would like to thank D. Greynat and S. Friot
for correspondence, and M.T. Seara in particular for discussions on
the subject of Ref.~\cite{Math}. DB, KM and SP would like to thank
the Department of Physics and Astronomy at San Francisco State
University for hospitality.  The work of D.B. is supported by the S\~ao Paulo
Research Foundation
(Fapesp) grant No. 2015/20689-9 and by the Brazilian National Council
for Scientific and Technological Development (CNPq), grant No.
305431/2015-3.
The work of IC was supported by the
Ministry of Research and Innovation of Romania, Contract PN 16420101/2016.
This material is based upon work supported by the U.S. Department
of Energy, Office of Science, Office of High Energy Physics, under
Award Number DE-SC0013682 (MG). KM is supported by a grant
from the Natural Sciences and Engineering Research Council of Canada,
and SP by
CICYTFEDER-FPA2014-55613-P, 2014-SGR-1450 and the
CERCA Program/Generalitat de Catalunya.

\section*{\Large APPENDICES}

\appendix
\begin{boldmath}
\section{Proof of Eq.~(\ref{bpt})}
\lbl{dirichletapp}
\end{boldmath}
Let us take the following Dirichlet series, corresponding to the
choice $\nu=1$ in $\Phi_1(s)$ in Eq.~(\ref{bigsplit}):
\be
\lbl{example1}
\chi(\sigma)= \sigma \sum_{n>n^*}^\infty \frac{\re^{-\sigma n}}{\log n} \ ,
\ee
with $n^*\gg 1$. The generalization to powers $\nu>1$ of the logarithm
is straightforward (see also App.~\ref{newapp} below). The function
$\chi(\s)$ is singular at $\s=0$ and we wish to find out what leading
the behavior of this function is as $\s \to 0^+$. We rewrite
\bea
\lbl{example2}
\chi(\sigma)&=&\sigma \int_0^\infty dt\ \sum_{n>n^*}^\infty
\re^{-\sigma n} \re^{-t \log n}= \sigma \int_0^\infty dt
\sum_{n>n^*}^\infty \re^{-\sigma n}\,  n^{-t} \\
&=&  \s \int_0^\infty dt\ \frac{1}{2 i \pi}\int_C ds \,
\sigma^{-s}\, \Gamma(s) \sum_{n>n^*}^\infty n^{-s-t} \nn \\
&=&\s \int_0^\infty dt\ \frac{1}{2 i \pi}\int_C ds \,
\sigma^{-s}\, \Gamma(s)\,  \zeta_>(s+t)\ ,\nn
\eea
where we have defined $\zeta_>(s+t)=\sum_{n>n^*} n^{-s-t}$. The
singular expansion of this function is the same as that of
$\zeta(s+t)$, namely $\zeta_>(s+t) \asymp 1/(s+t-1)$, and this
requires $C$ in Eq.~(\ref{example2}) to be a vertical straight line
with $\mathrm{Re}\, s>1$ . Clearly $\zeta_>(z)=\zeta(z)-\zeta_<(z)$,
where $\zeta_<(z)=\sum_{n\leq n^*} n^{-z}$ is a regular function of
$z$ containing no singularities. One can then split
\be
\lbl{split1}
\chi(\s)=\chi(\zeta,t_<;\s)- \chi(\zeta_<,t_<;\s)+\chi(\zeta_>,t_>;\s)\ ,
\ee
where
\bea
\lbl{split2}
\chi(\zeta,t_<;\s)&=&\frac{1}{2i\pi}\int_C ds\ \s^{1-s} \ \Gamma(s)
\int_0^{1-\delta} \ dt\ \zeta(s+t)\\
\chi(\zeta_<,t_<;\s)&=&\frac{1}{2i\pi}\int_C ds\ \s^{1-s} \ \Gamma(s)
\int_0^{1-\delta} \ dt\ \zeta_<(s+t)\nn\\
\chi(\zeta_>,t_>;\s)&=&\frac{1}{2i\pi}\int_C ds\ \s^{1-s} \ \Gamma(s)
\int^\infty_{1-\delta} \ dt\ \zeta_>(s+t)\nn
\eea
for a certain parameter $\delta$, with $0<\delta\ll 1$. We can now evaluate
or bound each of these integrals in turn.

Since the function $\Gamma(s)$ has poles at non-positive integers
and is regular for $s=1-t$ in the interval $0\leq t \leq 1-\delta$,
one may use $\zeta(s+t)\asymp 1/(s+t-1)$ and the Converse Mapping
Theorem \cite{Flajolet} to write
\bea
\lbl{chi1}
\chi(\zeta,t_<;\s)&=&\int_0^{1-\delta}dt\ \s^t\ \Gamma(1-t)
+\sum_{k=0}^{\infty}\frac{(-1)^k}{k!}\,\s^{1+k}
\int_{0}^{1-\delta}dt\ \zeta(t-k)\\
&=&\int_0^{1-\delta}dt\ \re^{t\log\s}\ \Gamma(1-t)+ \mathcal{O}(\s)\nn \\
&=&\sum_{k=0}^\infty c_k \int_0^{1-\delta} dt\, t^k\,
\re^{t\log\sigma}+ \mathcal{O}(\s)\ ,\nn
\eea
where the $\Gamma(1-t)$ has been expanded in powers of $t$, with
 $c_k \to 1$ as $k\to \infty$. Next, we split each term in the sum over
$k$ into the difference of an integral between $0$ and $\infty$, and
an integral between $1-\delta$ and $\infty$, with each of these integrals
being convergent (recall that we take $\sigma\to 0^+$).  This yields an
asymptotic expansion for $\chi(\zeta,t_<;\s)$ in powers of $1/\log\s$.
Using the saddle point method, one finds that
\be
\int_{1-\delta}^\infty dt\, t^k\,  \re^{t\log\sigma}
\sim \mathcal{O}\left( \frac{\s^{1-\delta}}{\log \s}\,(1-\delta)^k \right)\ ,
\ee
and we thus arrive at
\be
\lbl{example3}
\chi(\zeta,t_<;\s)=c_0\, \frac{\Gamma(1)}{(-\log\sigma)}
+c_1\, \frac{\Gamma(2)}{(-\log\sigma)^2}
+ c_2\,\frac{\Gamma(3)}{(-\log\sigma)^3}
+ \dots+ \mathcal{O}\left(\s,\frac{\s^{1-\delta}}{\log \s}\right)\ .
\ee
Notice that $(-\log\sigma)>0$ when $\sigma\to 0^+$, so the series
(\ref{example3}) is \emph{not} alternating, and thus this expansion
is not Borel summable.

\begin{figure}[thb]
\vspace*{4ex}
\includegraphics*[width=14cm]{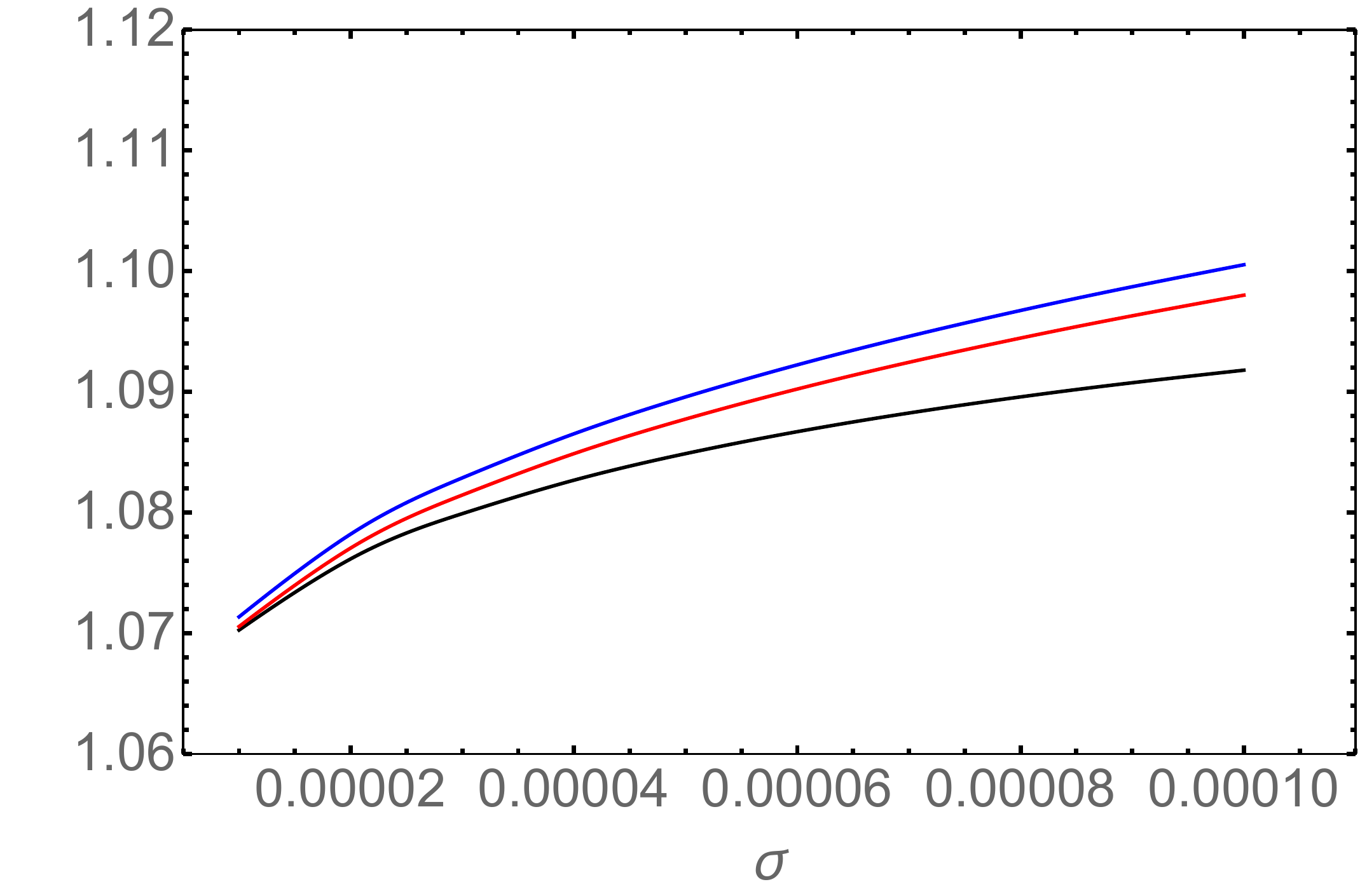}
\caption{Comparison of the Dirichlet series~(\ref{example1}) and
the result~(\ref{finalchi}). Blue curve: the Dirichlet
series~(\ref{example1}) with the sum running over $n$,
with $2\leq n < \infty$. Black curve: the same Dirichlet
series, but now for $20\leq n < \infty$. Red curve: the result
of the asymptotic expansion~(\ref{finalchi}) keeping 5 terms
in the series. The three functions have been multiplied by
$(-\log\sigma)$.}\label{}
\vspace*{-4ex}
\vspace{1cm}
\end{figure}

Let us now turn to the function $\chi(\zeta_<,t_<;\s)$ in
Eq. (\ref{split2}). The only singularities in the integrand
in  $s$ are those of $\Gamma(s)$. This leads to a result
of $\mathcal{O}(\s)$, which is already included in
(\ref{example3}).

Finally we have to evaluate $ \chi(\zeta_>,t_>;\s) $ in
Eq.~(\ref{split2}). Although originally the contour $C$
in this integral has $\mathrm{Re}\ s>1$, since the rightmost
singularity of $\zeta_>(s+t)$ is at $s=\delta$ in the interval
$1-\delta\leq t < \infty$, one is allowed to shift this contour
to the left to $\mathrm{Re}\ s=\delta'>\delta$ without changing
the result. Then, defining $s=\delta'+i u$ one finds
\be
\lbl{bound1}
 \chi(\zeta_>,t_>;\s) =\frac{\s^{1-\delta'}}{2\pi}
\int_{-\infty}^\infty du\ \s^{-i\, u}\ \Gamma(\delta'+i\, u)
\int_{1-\delta}^\infty dt\ \zeta_>(\delta'+i\, u + t) \ .
\ee
This function is bounded, \textit{i.e.},
\bea
\lbl{bound2}
|\chi(\zeta_>,t_>;\s)| &\leq& \frac{\s^{1-\delta'}}{2\pi}
\ M(\delta',\delta)\\
M(\delta,\delta')&=&\int_{-\infty}^\infty du\
\left|\Gamma(\delta'+i\, u)\right|
\ \int_{1-\delta}^\infty dt\ \left|\zeta_>(\delta'+i\, u + t)\right| \ .\nn
\eea
The integral for $M(\delta,\delta')$ is finite for the allowed region
of the parameters $\delta$ and $\delta'$. Since we may choose $\delta$,
and thus $\delta'>\delta$ arbitrarily small, we see that
$\chi(\zeta_>,t_>;\s)$ is exponentially suppressed in comparison
with the terms in the series~(\ref{example3}) for $\sigma\to 0^+$.

Putting together all the above results we conclude that
\be
\lbl{finalchi}
\chi(\s)= c^{(0)} \frac{\Gamma(1)}{(-\log\sigma)}
+c^{(1)} \frac{\Gamma(2)}{(-\log\sigma)^2}
+ c^{(2)}\frac{\Gamma(3)}{(-\log\sigma)^3}+ \dots
+ \mathcal{O}\left(\s,\s^{1-\delta'},\frac{\s^{1-\delta}}{\log \s}\right)\ ,
\ee
where $\delta>0$ and $\delta'>\delta$ can be chosen arbitrarily small.

In Fig.~3 we show a comparison of the original Dirichlet
series~(\ref{example1}) evaluated numerically and the
result~(\ref{finalchi}), after multiplying by $(-\log \sigma)$.
The difference between the blue and the black curves is the
starting value of $n$ in the sum  (\ref{example1}). The agreement as
 $\sigma \to 0^+$ shows that
the dependence on $\sigma$ as $\sigma\to 0^+$ depends solely
on the asymptotic behavior for large $n$ in this sum and not
on the first terms for low $n$. As one can see, the
result~(\ref{finalchi}) reproduces rather well the behavior of
$\chi(\sigma)$ as $\sigma\to 0^+$, but the corrections of
$\mathcal{O}\left(\Gamma(k)/(-\log\sigma)^k\right)$ are significant.

\begin{boldmath}
\section{Contributions from terms of the form
$\frac{1}{n^{\lambda} \log^{\nu} n}$}
\lbl{newapp}
\end{boldmath}
In this appendix we consider terms of the form
$n^{-s}\,\frac{1}{n^{\lambda} \log^{\nu} n}$ with $\lambda$ a positive
integer and $\nu$ a non-negative integer, as they appear in
$\Phi_2(s)$, {\it cf.} Eq.~(\ref{bigsplit}). We find it convenient to
first take $\lambda$ derivatives of the Mellin transform
\bea
\lbl{Mellinderivatives}
\frac{d^\lambda}{d\s^\lambda}\bh &=^{\hspace{-0.23cm}\circ}&(-1)^\lambda
\int_{\hat{C}} \frac{ds}{2i\pi}\ \s^{-s-\lambda}\ \Gamma(s+\lambda)
\ \sum_{n>n^*}n^{-s}\ \left(\frac{1}{n^{\lambda}\log^{\nu} n}\right) \\
&= &  (-1)^\lambda \int_{\hat{C}} \frac{ds}{2i\pi}\ \s^{-s-\lambda}\
\Gamma(s+\lambda)  \int_0^\infty dt\ \frac{t^{\nu-1}}{\Gamma(\nu)}
\ \zeta_>(s+\lambda+t) \nn \\
&\asymp &  (-1)^\lambda \int_{\hat{C}} \frac{ds}{2i\pi}
\ \s^{-s-\lambda}\ \Gamma(s+\lambda)
\int_0^\infty dt\ \frac{t^{\nu-1}}{\Gamma(\nu)}
\ \frac{1}{s+\lambda+t-1} \nn\\
&= &  (-1)^\lambda \int_0^\infty dt\ \frac{t^{\nu-1}}{\Gamma(\nu)}
\ \Gamma(1-t) \s^{t-1}+ \mathcal{O}\left( \s^0\right)\ .\nn
\eea
Again, our manipulations are formal, but the same justification that
applied in the case of $\Phi_1(s)$ applies also here (see
App.~\ref{dirichletapp}). In particular, the factor $\Gamma(1-t)$
inside the last integral should be understood as a power series in $t$.

We now distinguish three cases depending on the value of $\nu$:
$\nu=0$, $\nu=1$, or $\nu>1$.

\begin{itemize}
\item \underline{$\nu=0$}
Using that
  \be
\lim_{\nu \to 0}\int_0^\infty dt\,\frac{t^{\nu+k -1}}
{\Gamma(\nu)}\,e^{t\log\sigma}=\delta_{k,0}\ ,
  \ee
one finds that
   \be
   \frac{d^\lambda}{d\s^\lambda}\bh =^{\hspace{-0.23cm}\circ}
\frac{(-1)^\lambda}{\s}+ \mathcal{O}\left( \s^0\right)
\ee
which leads to
   \be
   \lbl{a}
   \bh =^{\hspace{-0.23cm}\circ} (-1)^\lambda\
\frac{\s^{\lambda-1}}{\Gamma(\lambda)}\ \log \s
+\mathcal{P}_\lambda(\s)\ , \quad \s \rightarrow 0^+\ ,
   \ee
where $\mathcal{P}_\lambda(\s)$ is a polynomial in $\s$ of degree $\lambda$.
Upon integration this yields, for large $|q^2|$ with $\mbox{Re}\,q^2<0$,
   \be
   \lbl{b}
   \mathcal{A}(q^2)=^{\hspace{-0.23cm}\circ} -\
\frac{\lambda}{(q^2)^\lambda}\ \log(-q^2) \left(1+
\mathcal{O}\left(\frac{1}{\log(-q^2)}\right)\right)
+\mathcal{P}_{\lambda+1}\left(\frac{1}{q^2}\right)\ .
   \ee
The polynomial in $1/q^2$ just modifies the series in powers of
$1/q^2$ we already found after Eq.~(\ref{bl}).

\item \underline{$\nu=1$}
Rewriting Eq.~(\ref{Mellinderivatives}) as
  \be
  \frac{d^\lambda}{d\s^\lambda}\bh =^{\hspace{-0.23cm}\circ}
\frac{(-1)^\lambda}{\s} \int_0^\infty dt\ \Gamma(1-t) \
\re^{-t \log 1/\s}+ \mathcal{O}\left( \s^0\right)\ ,
\ee
one obtains the asymptotic expansion (as $\s\to 0^+$)
  \be
\frac{d^\lambda}{d\s^\lambda}\bh =^{\hspace{-0.23cm}\circ}
\frac{(-1)^\lambda}{\s}\left( \frac{c_0\Gamma(1)}{-\log\s}
+ \frac{c_1\Gamma(2)}{(-\log\s)^2}+ \frac{c_2\Gamma(3)}
{(-\log\s)^3}+\dots\right) + \mathcal{O}\left( \s^0\right)\ ,
   \ee
where the coefficients $c_k$ are defined in App.~\ref{dirichletapp}.
Integrating this $\lambda$ times with respect to $\s$ yields
   \be
   \lbl{Bnu1}
   \bh=^{\hspace{-0.23cm}\circ}\frac{(-1)^{\lambda-1}}
{\Gamma(\lambda)}\,\s^{\lambda-1}\,\log(-\log\s)\left(1+
   \mathcal{O}\left(\frac{1}{\log(-\log\s)\log\s}\right)
\right)+\mathcal{P}_\lambda(\s)\ .
   \ee
Using Eq.~(\ref{adler}) one then gets
  \be
  \lbl{Adlernu1}
   \mathcal{A}(q^2)=^{\hspace{-0.23cm}\circ} -\
\frac{\lambda}{(q^2)^\lambda} \ \log\log(-q^2)\left( 1
+ \mathcal{O}\left( \frac{1}{\log(-q^2)\  \log\log(-q^2)}\right) \right)
   +\mathcal{P}_{\lambda+1}\left(\frac{1}{q^2}\right)\ .
\ee

\item \underline{$\nu>1$}

In this case one may integrate Eq.~(\ref{Mellinderivatives}) directly
  \be
  \bh =^{\hspace{-0.23cm}\circ} (-1)^\lambda \ \s^{\lambda-1}
\int_0^\infty dt \ \frac{t^{\nu-1}}{\Gamma(\nu)}\
\frac{\Gamma(1-t)\ \Gamma(t)}{\Gamma(t+\lambda)}\ \s^t
+ \mathcal{P}_\lambda(\s)\ .
  \ee
Writing $\s^t=e^{t\log\s}$ and expanding the rest of the
integrand in powers of $t$, one finds
  \be
  \lbl{c}
  \bh=^{\hspace{-0.23cm}\circ} \frac{(-1)^\lambda}{(\nu-1)
\Gamma(\lambda)}\ \frac{\s^{\lambda-1}}{(-\log\s)^{\nu-1}}
\left(1 + \mathcal{O}\left(  \frac{1}{\log \s}\right) \right)
+\mathcal{P}_\lambda(\s)\ ,
  \ee
which yields, for the Adler function,
  \be
  \lbl{d}
  \mathcal{A}(q^2)=^{\hspace{-0.23cm}\circ} \
\frac{\lambda}{\nu-1}\ \frac{1}{(q^2)^\lambda\
\log^{\nu-1}(-q^2)}\ \left( 1+ \mathcal{O}
\left( \frac{1}{\log(-q^2)}\right) \right)
+\mathcal{P}_{\lambda+1}\left(\frac{1}{q^2}\right)\ .
  \ee
\end{itemize}

Note that the results in Eqs.~(\ref{a}), (\ref{b}) are nothing but
the corresponding expressions, Eqs.~(\ref{c}), (\ref{d}), with $\nu$ set
equal to $0$.  Furthermore, the results in Eqs.~(\ref{c}), (\ref{d})
can be obtained from the expressions originally found in
Eqs.~(\ref{masterB}), (\ref{masteradler}) by the replacement
$\nu \rightarrow - \nu$. In other words, all these results are
connected by analytic continuation in $\nu$. The only exception is the
case {$\nu=1$, where the singularity in Eqs.~(\ref{c}), (\ref{d})
prevents the continuation to $\nu=1$. The bottom line
is that these logarithmic corrections show a clear hierarchy in
$\lambda$ and $\nu$: the more suppressed (relative to
the asymptotic Regge behavior) the correction $\frac{1}{n^\lambda \log^\nu n}$
as $n\to \infty$, the more suppressed the corresponding contribution to
the Adler function as $-q^2\to \infty$.

\begin{boldmath}
\section{$\Pi_{\rm DV}(q^2)$ from a branch point in the $\sigma$ plane}
\lbl{app1}
\end{boldmath}
Let us parametrize a branch-point singularity of the function $\borel$ as
\be
 \lbl{branch1}
 \borel=\frac{a_0}{(\s-\sh)^{1+\gamma}}\ \left[ 1
+ \frac{a_{\log}}{\log(\s-\sh)}+ a_1 \left (\s -\sh \right)^{p_1}
+\dots\right]\ ,
 \ee
where $p_1>0$ and the singularity is located at $\s=\sh=\s_0\,
\re^{i \phi_0}$, where $\s_0=|\sh|$ is the distance of the branch point to
the origin, and $\phi_0=\arg \sh$ is the angle with the positive real axis.
Parametrizing the cut as $\s=\sh + x\, \re^{i \phi}$, where
$0\leq x< \infty$ and $0\leq \phi < 2\pi$, with $\phi=\phi_0+\epsilon$
to the left of the cut and $\phi=\phi_0+2\pi-\epsilon$ to the right of
the cut in the bottom panel of Fig.~2 ($\epsilon \to 0^+$), one
finds for the discontinuity across the cut
\bea
\lbl{disc}
&&\mathrm{Disc}\left\{\borel\right\}=  2 i \sin(\pi\gamma) \
\mathrm{e}^{-i (1+\gamma) (\phi_0+\pi)}\ \frac{a_0}{x^{1+\gamma}}\\
&&\hspace{0.1cm}\times \left[ 1+a_1 x^{p_1}\ \mathrm{e}^{i p_1\phi_0}
+ \frac{a_{\log}}{2\, i \sin(\pi\gamma)}\ \re^{i\gamma \pi}
\int_0^\infty dt \left(\re^{it\phi_0}-\re^{-i2\pi(1+\gamma)}
\re^{it(\phi_0+2\pi)}\right) x^t + \dots\right]\ ,\nn
\eea
with this expression being valid for $x<1$.   A similar expression
can be derived for $x>1$, but the contribution from the discontinuity
to Eq.~(\ref{result1}) turns out to be exponentially suppressed in $q^2$ relative to the
result shown in that equation, as can be shown using arguments similar to those
used in App.~\ref{dirichletapp}.

Then, using that
\be
\int_0^\infty dx\, \re^{-x a} x^b = \frac{\Gamma(1+b)}{a^{1+b}}\ ,
\ee
one obtains
\be
\lbl{result1}
\Gamma(1+\gamma) \int_{\Gamma}d\s\  \re^{\s q^2}\
\mathrm{Disc}\left\{\frac{1}{(\s-\sh)^{1+\gamma}}\right\}=
2\pi i \, \re^{-i \gamma \pi} \, (-q^2)^\gamma\, \re^{q^2\sh}  \ ,
\ee
where the identity~(\ref{identity}) has been used to bring
the result into a form where it is evident that the usual
residue theorem result is obtained in the limit $\gamma\to 0$.
An analogous calculation yields, for $q^2>0$,
\be
\int_{\Gamma}d\s\  \re^{\s q^2}\ \mathrm{Disc}\
\left\{\frac{1}{(\s-\sh)^{1+\gamma}\ \log(\s-\sh)}\right\}=
\frac{2\pi i \re^{-i\gamma\pi} (-q^2)^\gamma \re^{q^2\sh}}
{\Gamma(1+\gamma)\ \log q^2 }\left(1+\mathcal{O}
\left(\frac{1}{\log q^2}\right)\right)\ .
\ee
Gathering all the terms, one finally obtains for
Eq.~(\ref{DVbranch}) the expression
\bea
\lbl{resultNc1}
&& \Pi_{\rm DV}(q^2)=  2  \pi i\  \frac{\re^{-i \pi\gamma}}
{\Gamma(1+\gamma)} (-q^2)^\gamma\ \re^{ q^2 \sh}a_0 \\
&&\hspace{1cm}\times\left[1+ a_1\ \frac{ \Gamma(1+\gamma)}
{\Gamma(1+\gamma-p_1)}\ \frac{\re^{i\pi p_1}}{ (-q^2)^{p_1}}
+ \frac{a_{\log}}{\log q^2}\left(1+\mathcal{O}
\left(\frac{1}{\log q^2}\right)\right)  +\dots \right]\ .\nn
\eea

\vspace{0.4cm}
\begin{boldmath}
\section{Some numerical results}
\lbl{sec:numerics}
\end{boldmath}
Here we compare the results obtained from fits to hadronic
$\tau$-decay data in Ref.~\cite{Boito3}
to those obtained from fits to Regge trajectories
in Ref.~\cite{Masjuan}.

For the latter, Ref.~\cite{Masjuan} finds, from fits of the meson
spectrum to radial trajectories, the value
\be
\lbl{lambda}
\Lambda^2\simeq 1.35(4)\ \mathrm{GeV}^2\ .
\ee
for the slope of these trajectories, and, from an average over
light-quark meson states, the value
\be
\lbl{rho}
\frac{\Gamma}{M}\simeq 0.12(8) \simeq \frac{a}{N_c}
\ee
for the angle, $\varphi_{N_c}$.
For comparison, for the $\rho$, this ratio is equal to approximately
$0.19$.

These results are to be compared to those obtained in Ref.~\cite{Boito3}
from finite-energy sum-rule fits to variously weighted integrals of
hadronic $\tau$-decay data, in which para-metrizations of the form
$\rho_{\rm DV}(t) \propto \mathrm{e}^{-\gamma t}\sin(\alpha+\beta t)$
were employed for the duality violating parts of the vector and axial
vector current spectral function at large $t$ ($t>s_{min}$). The results
of the fits to the non-strange, vector channel data for the parameters
$\beta_V$ and $\gamma_V$, are
\be
\lbl{fits}
\beta_V= 4.2(5)\ \mathrm{GeV}^{-2}\ , \qquad
\gamma_V = 0.7(3)\ \mathrm{GeV}^{-2}\ ,
\ee
where the errors include variations of the results over the different fit
types (one- or three-weight fits) and $s_{min}$ ranges explored in
Ref.~\cite{Boito3}.   These two numbers are seen to agree well with the
result in Eq.~(\ref{final result}), which,  after re-introducing physical
units and together with Eqs.~(\ref{lambda})
and~(\ref{rho}), yields
\be
\lbl{numerics}
\beta_V =\frac{2\pi}{\Lambda^2}\simeq 4.7(2)\ \mathrm{GeV}^{-2}
\ ,\qquad \gamma_V =\frac{2\pi}{\Lambda^2}\,\frac{a}{N_c} \simeq 0.6(4)\
\mathrm{GeV}^{-2}\ .
\ee
The factors of $2\pi$ in Eq.~(\ref{numerics}), are crucial for the
agreement with Eq.~(\ref{fits}). The agreement thus goes well
beyond that of a simple estimate based solely on naive dimensional
analysis.

\vspace{0.4cm}

\end{document}